\documentclass[12pt]{article}

\usepackage{epsfig}

\def\bb{\mbox{\bf b}}

\def\bs{\mbox{\bf s}}
\def\bt{\mbox{\bf t}}
\def\bc{\mbox{\bf c}}

\def\bphi{\mbox{\boldmath $\phi$}}

\def\balpha{\mbox{\boldmath $\alpha$}}
\def\bbeta{\mbox{\boldmath $\beta$}}
\def\btau{\mbox{\boldmath $\tau$}}
\def\bnu{\mbox{\boldmath $\nu$}}

\def\bPhi{\mbox{\boldmath $\Phi$}}

 \def\tfrac#1#2{{\textstyle{{#1}\over{#2}}}}

\def\half{\tfrac{1}{2}}
\def\third{\tfrac{1}{3}}
\def\quarter{\tfrac{1}{4}}

\def\twothirds{\tfrac{2}{3}}

\def\tr{\mathop{\rm tr}\nolimits}

\usepackage{amssymb}

\def\bbz{{\mathbb Z}}

\def\br{{\rm BR}}

\begin{document}

\begin{titlepage}

\baselineskip 24pt

\begin{center}

{\bf {\Large Two variations on the theme of Yang and Mills---the 
SM and the FSM}}

\vspace{.5cm}

\baselineskip 14pt

{\large CHAN Hong-Mo}\\
hong-mo.chan\,@\,stfc.ac.uk \\
{\it Rutherford Appleton Laboratory,\\
  Chilton, Didcot, Oxon, OX11 0QX, United Kingdom}\\
\vspace{.2cm}
{\large TSOU Sheung Tsun}
\footnote{and virtually also Jos\'e Bordes, with whom almost 
all the reported work on the FSM was done, although he has not 
taken part in the actual writing of the present article.}\\
tsou\,@\,maths.ox.ac.uk\\
{\it Mathematical Institute, University of Oxford,\\
Radcliffe Observatory Quarter, Woodstock Road, \\
Oxford, OX2 6GG, United Kingdom}\\

\end{center}

\vspace{.3cm}

\begin{abstract}

The standard model (SM) is viewed as a variation on the Yang-Mills 
theory with gauge symmetry $u(1) \times su(2) \times su(3)$, in 
which the flavour symmetry is framed and to which 3 generations of 
quarks and leptons are appended as inputs from experiment.  The 
framed standard model (FSM) is then a further variation on the SM 
in which the colour symmetry is also framed, where the 3 generations 
now follow as consequences together with their characteristic mass 
and mixing patterns.  In addition, the FSM yields as a bonus a 
solution to the strong CP problem leading to a unified treatment 
seemingly of all known CP physics for both quarks and leptons.  It 
predicts, however, also a ``hidden'' sector populated by particles 
yet unknown, hence full both of promises (e.g.\ dark matter?) and of 
threats, which is just beginning to be explored (i) with a modified 
Weinberg mixing and (ii) with the $g-2$ and some other anomalies.

\end{abstract}

\end{titlepage}

\newpage

The Yang-Mills theory \cite{ym} is an extension of the abelian gauge 
theory of electromagnetism to nonabelian symmetries.  Although when 
first proposed in 1954, it was not exactly clear to what physical 
situation it applied, it soon became in the decades that followed 
the basis of almost every serious contender for the fundamental 
theory of the physical world.  

\section{The standard model as a variation of Yang-Mills}

In particular the standard model (SM) of particle physics has for 
its backbone a Yang-Mills theory with gauge symmetry 
\begin{equation}
G = u(1) \times su(2) \times su(3).
\label{G}
\end{equation}  
And this standard model can justly claim to be the most successful 
physical theory ever in terms of the wide range of phenomena it 
covers and the accuracy with which it does so.  It has survived 
all the intense scrutiny it has been subjected to by experiment with 
flying colours, and only recently have there been reported some 
possible small deviations \cite{g-2exptmuon,g-2Fermilab,Bdecay} from what it 
prescribes.

The SM, however, is not just an application of the Yang-Mills idea 
to the special case $u(1) \times su(2) \times su(3)$.  It is rather
a variation on the Yang-Mills theme in that to the basic Y-M gauge 
structure is added a Higgs scalar field $h_W$.  This is essential 
for breaking the flavour $su(2)$ symmetry and for giving masses to 
the particles observed.  While stressing the significance of the 
Y-M structure, therefore, that of this important variation should 
not be ignored, for it is only with the addition of this latter 
that the SM is enabled to make contact with reality and achieve its 
phenomenal success.

\section{'t~Hooft's confinement picture of electroweak}

The effect of the Higgs field in giving masses to particles in the 
electroweak theory is usually pictured as a spontaneous breaking of
the flavour gauge symmetry but, as pointed out by 't~Hooft in an 
insightful paper \cite{tHooft}, it can equally be pictured as a 
theory in which local flavour $su(2)$ is confining and exact; what 
is broken is only a global symmetry hidden in the theory, say 
$\widetilde{su}(2)$, which is associated or dual to it.  The massive 
particles that are observed in experiment, such as the Higgs boson 
$h_W$ and the vector bosons $Z, W$ as well as the quarks and 
leptons, would appear in this confinement picture as bound states by 
flavour confinement of the fundamental scalar with respectively its 
conjugate (in $s$-wave as $h_W$, in $p$-wave as $Z, W$) and 
with the fundamental fermion field (as quarks and leptons).  This is 
an alternative picture that some, including ourselves, may sometimes 
find easier to envisage and will be adopted for our later discussions.

\section{Fundamentalist's questions on the SM}

Despite its great success in confrontation with data, however, the 
SM faces some questions from the fundametalist, in particular, the 
following:

\begin{itemize}

\item {\bf [Q1]} Why the Higgs scalar field?

In the original Y-M framework, the vector bosons appear naturally, 
say as connections in a fibre bundle, which is of course part of 
its attraction.  To this, however, the SM, so as to match experiment, 
has now added the scalar field as something of a phenomemological 
afterthought which seems to have spoiled the purity of the original 
picture unless, one feels, a geometrical meaning be found also for 
the scalar field.

\item {\bf [Q2]} Why 3 generations each of quarks and leptons, and 
why in such mass and mixing patterns?

The original Y-M framework admits any number of fermions but, again 
to satisfy experimental demand, the SM postulates 3 (and apparently 
only 3) copies each of both quarks and leptons, called generations, 
without giving any reason for their occurence.  Further, these quarks 
and leptons are assigned different masses and required to mix with 
one another in particular patterns, with mass and mixing parameters 
having widely different values taken just as inputs from experiment, 
and left by the SM unexplained.\footnote{This so-called ``generation 
problem'' is in fact the question that Weinberg stated in
an interview at CERN \cite{Weinbergint} as the one that he would like 
to know the answer to before he dies. It is also the question to which 
he has devoted one of his last published research papers 
\cite{Weinberggp}.  Sadly, with his passing, we shall now never know 
what his answer to this question would be.}  Together, they account 
for about two-thirds of the SM's some thirty parameters.

\end{itemize}

In other words, both the Higgs field and the 3 generations of quarks 
and leptons are injections from phenomenology which have detracted 
from the theoretical purity and appealing beauty of the original Y-M 
framework.  And the fundamentalist would like to see a theoretical 
understanding of both these injections to have the original purity 
of the Y-M theory restored.

Now these questions are not a criticism of the SM, which as far as 
we know is consistent within itself and also with the data it sets 
out to describe.  But they do suggest that there is perhaps some
structure deeper behind the SM that we do not yet understand, of 
which the SM is but the surface manifestion.  To these questions, 
of course, many physicists, including ourselves, have devoted a lot 
of energy and we would like to tell you in what follows some of 
the answers we have so far come across in a scheme which goes by 
the name of the framed standard model (FSM).

\section{The Higgs field as frame vector in flavour space}

Focussing first on {\bf [Q1]}, we noted many years back \cite{mbsm}, 
as others might have done before us, that the Higgs scalar had the 
same transformation properties as frame vectors in flavour $su(2)$ 
space.  Explicitly, the Yang-Mills theory for $su(2)$ is built to 
be invariant under local gauge transformations representable as 
$2 \times 2$ matrices, say:
\begin{equation}
\Phi = (\phi_r^{\tilde{r}}), \ \ r = 1, 2; 
   \ \ \tilde{r} = \tilde{1}, \tilde{2},
\label{Phi}
\end{equation}
transforming from the local (i.e.\ space-time point $x$-dependent) 
frame indexed by $r$ to a global (fixed, $x$-independent) reference 
frame indexed by $\tilde{r}$.  The two columns of $\Phi$, say the
``frame vectors'' $\bphi^{\tilde{r}}$, transform as doublets under 
flavour $su(2)$ but as scalars under proper Lorentz transformations, 
in other words, the same as the Higgs field in the electroweak (EW) 
theory.  

Promoting the frame vectors $\bphi^{\tilde{r}}$ into fields to be 
dynamical variables in the $su(2)$ Yang-Mills theory, or ``framing 
the $su(2)$ theory'' for short, is thus similar in spirit to taking 
the vierbeins as dynamical variables in gravity, although 
of course vierbeins are frame vectors in ordinary spacetime while 
$\bphi^{\tilde{r}}$ are frame vectors in internal symmetry space.

There are 2 frame vectors $\bphi^{\tilde{r}}$, however, while in 
the standard electroweak theory, there is only 1 Higgs field.  One 
can conform to this by insisting that the frame vector fields (or 
``framon fields'' for short) $\bphi^{\tilde{r}}$ should satisfy the 
condition:
\begin{equation}
\phi^{\tilde{2}}_r = - \epsilon_{rs} (\phi^{\tilde{1}}_s)^*,
\label{minframe}
\end{equation}
which the frame vectors originally satisfy, and allow one of the 
framon fields, say $\bphi^{\tilde{2}}$, to be eliminated in terms 
of the other, say $\bphi^{\tilde{1}}$, thus minimizing the number 
of new dynamical variables to be introduced by framing.  

One can regard thus the standard electroweak theory as a minimally 
framed variation of Yang-Mills and restore somewhat the geometric 
purity of conception in the original.  There is also one slight 
advantage in this framonic viewpoint in that the ``hidden'' global 
symmetry noted in the usual formulation of the electroweak theory, 
referred to as $\widetilde{su}(2)$ in Section 2, is here automatic, 
being built into the theory already when the Higgs field is taken 
as a frame vector, since an $\widetilde{su}(2)$ transformation on 
$\tilde{r}$ indices means just a change of basis in the reference 
frame, under which physics should of course be invariant.  Hence:
\begin{itemize}
\item {\bf [R1]}  The Higgs field in the EW theory is given a 
hitherto missing geometric meaning as a frame vector field,
\end{itemize}
restoring thus, to some extent, the theoretical purity of the 
original Y-M scheme, i.e. as far as the EW theory is concerned.

\section{What about framing colour?}

In our present language then, the SM and all phenomenal successes 
it has achieved so far have arisen by framing the flavour theory. 
But if framing is of such significance, why should it be applied 
only to the flavour but not to the colour component of the gauge 
symmetry $G$ in (\ref{G}) as well?  At first sight, the answer 
might seem obvious: the colour theory is conceived as confining 
so that the colour symmetry has to remain exact, while flavour is 
generally conceived as being broken, and the Higgs field (or in 
our present language ``framon field'') was introduced explicitly to 
achieve that end.  

However, if one takes account of 't~Hooft's confinement picture 
of the flavour theory as outlined in Section 2, the basis of 
this difference in treatment between flavour and colour becomes 
unclear.  It is seen there, in the flavour theory, that a local 
symmetry being confining and exact is not inconsistent by itself 
with framing, since the broken symmetry can be taken as the dual 
global symmetry instead.  The same can result in principle in 
framing the colour theory also.  In that case, we can retain 
colour confinement as experiment indicates while gaining a new
broken global symmetry, say $\widetilde{su}(3)$.  Explicitly,
suppose  we introduce for the colour theory, in parallel to 
(\ref{Phi}) in Section 4:
\begin{equation}
\bPhi = (\phi_a^{\tilde{a}}), \ \ a = 1, 2, 3; 
   \ \ \tilde{a} = \tilde{1}, \tilde{2}, \tilde{3},
\label{bPhi}
\end{equation}
there is indeed a global $\widetilde{su}(3)$ symmetry built-in 
in the system corresponding to basis changes in the reference 
frame labelled by the tildered indices.  

Phenomenologically, this new global $\widetilde{su}(3)$ symmetry
would be no bad thing to gain, for it could conceivably play the 
role of the long-sought-after symmetry for fermion generations 
\cite{tfsm}.
In other words, when viewed in this way, framing the colour 
theory is not only admissable but might even lead to a solution 
of the generation problem, or in other words, an answer to the 
question {\bf [Q2]} of Section 3 that Weinberg wanted.

\section{Framing the standard model}

That being the case, the temptation is there to consider framing 
as of a sort of new physics principle to be applied to all gauge 
theories.   This would restore to some extent the theoretical 
purity that the Y-M theory originally possessed while offering a 
chance, as the last section suggests, of answering the question 
{\bf [Q2]} posed above on the SM.  With this observation in mind, 
let us go back to the beginning with SM first as a Yang-Mills 
theory with local gauge symmetry $G$ in (\ref{G}), but instead of 
framing just the flavour $su(2)$ factor as the usual SM did, we 
push now the framing idea all the way to include all of $G$.  
What will then result?  

Framons we recall are frame vectors and carry two sets of indices, 
one set referring to the local frame, and the other referring to 
the global reference frame, and physics should be invariant under 
basis changes in both the local and the global reference frames.  
This means that the fully framed standard model (FSM) including 
the framons as dynamical variables should be invariant under both 
the local symmetry:
\begin{equation}
G = u(1) \times su(2) \times su(3)
\label{G}
\end{equation}
and its ``dual'' global symmetry:
\begin{equation}
\tilde{G} = \tilde{u}(1) \times \widetilde{su}(2) \times 
   \widetilde{su}(3).  
\label{Gtilde}
\end{equation}

And the framons themselves are to be representations of both $G$ 
and $\tilde{G}$, but are scalars under Lorentz transformations.
It was suggested already in (\ref{Phi}) and (\ref{bPhi}) above 
that they should be doublets ${\bf 2}$ in flavour $su(2)$ and 
triplets ${\bf 3}$ in colour $su(3)$ and carry presumably also a 
$u(1)$ charge $y$ or ${\bf 1}$.  However, $G$ being a product of 
these 3 factor symmetries, there is a choice of the framon being 
assigned the sum or the product representation of any pair of the 
3 factors symmetries, namely ${\bf 1} (+/\times) {\bf 2} (+/\times)
{\bf 3}$.  After some trial and error considering the possible 
physical consequences of each, one comes to the conclusion that 
the most suitable choice is:
\begin{equation} 
{\bf 1} \times ({\bf 2} + {\bf 3}).
\label{minframeG}
\end{equation}
This choice happens also to introduce the smallest number of new 
scalar fields, so that one can think of the FSM as a minimally 
framed theory with gauge symmetry $G$ in the same spirit as the 
standard EW theory was said in section 4 to be a minimally framed 
Y-M gauge theory with symmetry $u(1) \times su(2)$. 

Explicitly, this gives the framons then as:
\begin{itemize}
\item (FF) the flavour (``weak'') framon:
\begin{equation}
\balpha \bphi = \left( \alpha^{\tilde{a}} \phi_r \right);
    \ \ r = 1, 2; \ \  \tilde{a} = \tilde{1}, \tilde{2},
    \tilde{3},
\label{fframon}
\end{equation}
with $q = \pm \half$.
\item (CF) the colour (``strong'') framon:
\begin{equation}
\bbeta \bPhi = \left( \beta^{\tilde{r}} \phi_a^{\tilde{a}}
    \right); \ \  a = 1, 2, 3; \ \  \tilde{a} = \tilde{1},
    \tilde{2}, \tilde{3}; \ \  \tilde{r} = \tilde{1},
    \tilde{2}.
\label{cframon}
\end{equation}
with $q = - \third, + \twothirds$,
\end{itemize}
where $\balpha (\bbeta)$ is a global colour triplet (flavour 
doublet) which, for here, can be taken as a real unit 3-vector 
in dual colour (dual flavour) space.

\section{The FSM action}

Augmenting the SM action with the introduction of these framons as 
dynamical variables in addition to the usual gauge vector boson 
and matter fermion fields then completes the specification of the 
FSM \cite{tfsm}.   In addition to the SM action involving only the 
gauge boson and matter fermion fields, there are now 3 extra terms 
involving the framons as well: (i) the framon self-interaction 
potential $V[\Phi]$, (ii) the ``framon kinetic energy'' term 
coupling the framon to the gauge bosons, and (iii) the ``Yukawa 
terms'' coupling the framons to the matter fermion fields.  All 3 
terms have to be invariant under the doubled symmetry $G \times 
\tilde{G}$, which requirement restricts quite stringently the form 
they each could take.  For example, the framon potential turns out 
to have the form: 
\begin{eqnarray}
V & = & - \mu_W |\bphi|^2 + \lambda_W (|\bphi|^2)^2 \nonumber \\
  &   & - \mu_S \sum_{\tilde{a}} |\bphi^{\tilde{a}}|^2
        + \lambda_S \left( \sum_{\tilde{a}} |\bphi^{\tilde{a}}|^2 \right)^2
        + \kappa_S \sum_{\tilde{a} \tilde{b}} |\bphi^{\tilde{a}*}.\bphi^{\tilde{b}}|^2 
          \nonumber \\
  &   & + \nu_1 |\bphi|^2 \sum_{\tilde{a}}|\bphi^{\tilde{a}}|^2  
        - \nu_2 |\bphi|^2 \left( \sum_{\tilde{a}} \alpha^{\tilde{a}} 
          \bphi^{\tilde{a}}\right)^\dagger 
          \cdot \left( \sum_{\tilde{a}} \alpha^{\tilde{a}} \bphi^{\tilde{a}} \right),
\label{V}
\end{eqnarray}
(up to quartic terms for renormalizability) depending on 7 real
coupling parameters.  The explicit invariant forms of the other 
2 terms are given in \cite{tfsm,cfsm} but will not be needed in 
this paper.  Together with the SM action, these new added terms 
define then the FSM.

This FSM will be a variation on the SM theme as the latter is a 
variation of Y-M, or a second variation on the Y-M theme itself.
As suggested above, it has the hope of:
\begin{itemize}
\item Recovering for the SM variation the theoretical purity of 
the original Y-M theme by assigning to the Higgs scalar and the 
3 fermion generations each a basic (geometric) significance.
\item Reducing the SM's reliance on experimental input, i.e.\ reducing 
the number of empirical parameters the SM has to introduce.
\end{itemize}
Besides, being a new theory with many new degrees of freedom in 
terms of the colour framon, it offers also the hope of:
\begin{itemize}
\item Possibly opening up vistas of fundamentally new physics 
for exploration.
\end{itemize}
Whether these hopes will be realized to some extent, or will be 
dashed early by implications irreconcilable with experiment can 
only be answered by working out the details.

\section{The FSM vacuum}

To do so, one needs first to work out the FSM vacuum by minimizing 
the framon potential $V$ in (\ref{V}).  This was done in, for 
example, \cite{tfsm}.  The flavour vacuum is essentially the 
same as in the standard EW theory but the colour vacuum is more 
intricate.  The latter depends on $\balpha$ from the flavour 
framon {\bf[FF]} because of the $\nu_2$ term in (\ref{V}) linking 
{\bf [FF]} to the colour framon {\bf [CF]} and also, of course 
on the gauge choice.  In the local colour $su(3)$ gauge where 
$\bPhi$ is hermitian and the global colour $\widetilde{su}(3)$ 
gauge where $\balpha$ points in the third direction, the colour 
vacuum turns out to have the following form:
\begin{eqnarray}
\balpha_{\rm vac}^\dagger & = & (0, 0, 1), \\
\bPhi_{\rm vac} & = & \zeta_S \left( \begin{array}{ccc} Q & 0 & 0 \\
                                                     0 & Q & 0 \\
                                                     0 & 0 & P 
                           \end{array} \right),
\label{Phivac}
\end{eqnarray}
with:
\begin{equation}
Q = \sqrt{\third (1 - R)}, \ \ P = \sqrt{\third(1 + 2R)}, \ \ 
R = \frac{\nu_2 \zeta_W^2}{2 \kappa_S \zeta_S^2}.
\label{PQR}
\end{equation}
We note that the dual colour symmetry which is to play the role 
of generations for fermions is broken as expected.

\section{Particle masses, couplings, and flavour-colour dichotomy}

Substituting the above vacuum into the action and expanding about 
it, one obtains the mass matrices and couplings of the particles 
in the theory at tree level.  This was done in \cite{cfsm}.  One 
finds that:

\begin{itemize}
\item The particles which appear separate naturally into 2 sectors:
\begin{itemize}
\item {\bf [SS]} comprising the particles we know, namely the Higgs 
boson $h_W$, the vector bosons $W, Z$, and the quarks and leptons, 
appearing as bound states via flavour confinement of the flavour 
framon with its own conjugate or with the fundamental fermions, as 
explained in section 2.
\item {\bf [HS]} comprising new particles unfamiliar to us, 
which appear as bound states via colour confinement of the colour 
framon first with its own conjugate in $s$-wave labelled generically 
as $H$, and in $p$-wave labelled generically as $G$, then secondly 
with fundamental fermion fields, labelled generically as $F$.  They 
are the exact parallels of respectively the $h_W$, $W-Z$, and quarks 
and leptons in {\bf [SS]} above.
\end{itemize}
except for the photon which stands on its own and couples to both 
sectors 
\footnote{and possibly some particles belonging to both sectors 
formed each as a three-body bound state from (i) a fundamental 
fermion carrying both flavour and colour with (ii) a flavour framon 
via flavour confinement and (iii) a colour framon via colour 
confinement.  We are unsure whether such states exist because they 
cannot be handled with the current tools used, but if they do, they 
may play a crucial physical role in linking the two sectors together.}.
\item The photon mixes at tree-level with the $Z$ in {\bf [SS]} and 
with a particle called $G_8$ in {\bf [HS]}, while the Higgs boson 
$h_W$ in {\bf [SS]} mixes with 2 particles called $H_+$ and $H_3$ in 
{\bf [HS]}.
\item Apart from these mixings, there is no coupling linking particles 
from {\bf [SS]} to particles from {\bf [HS]} although, of course, 
there are many intricate couplings among particles within the same 
sector which can lead to further structures.
\item For example, in the standard sector, we know that the quarks, 
formed as framon-fermion bound states via flavour confinement, still 
carry colour and will combine with one another via colour confinement
to form nucleons, which in turn will combine via soft nuclear forces 
to form nuclei etc.  So in the hidden sector in parallel, some $F$s, 
though formed themselves as framon-fermion bound states via colour 
confinement, may still carry flavour, which we can call co-quarks.  
And these co-quarks can further combine with one another via flavour 
confinement to form co-hadrons etc. 
\item However, the two sectors can communicate only via the exchange 
of photons or via the mixings of the portal states $Z,\;h_W$ on one 
side and $G_8, H_+, H_3$ on the other.  In other words, the lowest states
in {\bf [HS]} if neutral in eletric charge would seem ``dark'' to us.  
This lack of communication can make the sector {\bf [HS]} hard to 
access for us living in the standard sector {\bf [SS]} and suggests 
the label HS (hidden sector) for it.
\end{itemize}

That the particle spectrum in FSM should be separated into two 
sectors with the roles of flavour and colour interchanged (which we 
shall call the ``flavour-colour dichotomy'' for short) is perhaps no 
surprise, at least with hindsight, given that in its formulation the 
FSM has treated flavour and colour even-handedly and that the sum 
representation ${\bf 2 + 3}$ has been chosen for the framon.  This 
does not mean, however, that the two sectors will have all properties
similar for there are sufficient differences between the flavour 
and colour symmetries themselves to lead in the two sectors to very 
different results.  That we ourselves should live in one sector, the 
{\bf [SS]}, and have thereby difficulty communicating with the other 
sector {\bf [HS]} means that, to us, the other sector is ``hidden'' 
and may be to some extent ``dark''.  But in actual fact, given the 
complexity of the particle spectrum and of the couplings among the 
particles as seen in \cite{cfsm}, the ``hidden sector'' may in fact 
be even more vibrant within itself than our own.  Indeed, even in 
degrees of freedom, the colour framon which populates {\bf [HS]} has 
9 compared with the flavour framon which populates our sector having 
only 2, meaning that of the two sectors, it is the {\bf [HS]} which 
will be the more populous.\footnote{It is amusing to note that the 
ratio 2:11(=9+2) is tantalizingly close to the observed fraction of 
around 18 percent of lunminous to all matter \cite{pdg}.}

The possible existence of a ``hidden sector'' {\bf [HS]} yet unknown 
to us is likely to become the FSM prediction which will stretch most 
our imagination and its credibility.  But we shall leave indulgence 
in the excitement of its exploration until later (Sections 12, 13)
and concentrate first on the practical, dealing with the standard 
sector {\bf [SS]} comprising the particles we already know.

\section{Consequence: 3 fermion generations with characteristic mass 
and mixing patterns}

Our first attention was naturally the quarks and leptons in {\bf [SS]} 
the standard sector which were the original motivation for constructing 
the FSM.  This was studied in \cite{tfsm}.  We briefly summarize here 
the result. 

The mass matrix at tree level of all quarks and leptons, obtained by 
substituting in the Yukawa coupling for the flavour framon its vacuum 
expectation value, turns out to have the following simple form:
\begin{equation}
m = m_T \balpha \balpha^\dagger,
\label{mfact}
\end{equation}
where $\balpha$ comes from the flavour framon as seen in {\bf [FF]} 
and is independent of the fermion type (or species), i.e.\ whether up-
or down-type quarks, or charged lepton or neutrinos.  
The numerical coeffcient 
$m_T$, however, depends on the fermion type.  We recall that $\balpha$ 
is a 3-vector in $\widetilde {su}(3)$, and $m$ a $3 \times 3$ matrix. 
Hence, it follows that 
\begin{itemize}
\item {\bf [R2]} all quarks and leptons occur in 3 generations,
\end{itemize}
answering partly the question {\bf [Q2]} posed above in section 3.  

This mass matrix has only 1 massive state and zero mixing, which is 
not bad for a tree-level approximation, at least for quarks where 
$m_t >> m_c, m_u, m_b >> m_s, m_d$ and the off-diagonal elements are
small in the CKM matrix.

One can do better, however, since the knowledge of the tree-level mass 
and coupling parameters obtained before allows one to carry calculations 
to loop-levels.  This was done in \cite{tfsm} to 1-framon-loop, and the 
result is that the vector $\balpha$ rotates with changing scales but the 
form of the mass matrix ({\ref{mfact}) remains the same.  The reason for 
this rotation is that the framon, alone among the fundamental fields in 
the FSM, carry both the local $su(3)$ and global $\widetilde{su}(3)$ 
indices, so that framon loops can change the relative orientation 
between the local and global frames.  This is also the reason why the 
emphasis in \cite{tfsm} was put on framon loops rather than other loops 
which change only the normalization.  

And once $\balpha$ rotates with scale, the degeneracy of the tree-level 
mass matrix is broken to admit both non-zero lower generation masses 
and mixings.  That this attains is an idea with a rather long history 
much predating the FSM, although worked on, as far as we know, only by 
Bjorken \cite{Bj} and by our group \cite{r2m2}.  The point is that for 
a rotating rank-1 mass matrix (R2M2) such as (\ref{mfact}), both the 
eigenvalues and eigenvectors depend on the scale so that the masses 
and state vectors of the physical states can no longer be obtained just 
by diagonalizing the mass matrix at any one scale.  Inputting then the 
concept that the physical masses and state vectors are to be determined 
for each particle at its own mass scale, one comes to the conclusions:
\begin{itemize} 
\item {\bf [qf1]} 
that there will be mixing between up and down states.  Take for 
example the $t$ and $b$ quarks.  The state vector $\bt$ of $t$ is given 
in (\ref{mfact}) as $\balpha(\mu = m_t)$, and that of $b$ similarly as 
$\bb = \balpha(\mu = m_b)$.  Since $m_t \neq m_b$ and $\balpha$ rotates 
by assumption, $\bt \neq \bb$, and hence $V_{tb} = \bt \cdot \bb \neq 1$.
In other words, there is mixing as claimed.
\item {\bf [qf2]}
that lower generations will acquire nonzero masses by ``leakage''.  
Take for example the $c$ quark.  Being an independent quantum state to 
$t$, its state vector $\bc$ must be orthogonal to $\bt$, i.e.\ to $\balpha$ 
at $\mu = m_t$.  But since $\balpha$ rotates with $\mu$ by assumption, 
$\balpha(\mu = m_c)$, which continues to carry the mass, will now have a 
nonzero component in the direction of $\bc$, giving thus $c$ a mass by 
``leakage''.
\item {\bf [qf3]}
what may take a little more to figure out, that the corner elements 
of the mixing matrices ($V_{ub}, V_{td}$ in CKM and $U_{e3}$ in PMNS)  
are associated with the twist (torsion) of the $\balpha$ trajectory and 
are therefore smaller in magnitude than the other mixing elements 
($V_{us},V_{cd},V_{cb},V_{ts}$ in CKM, $U_{e2}, U_{\mu 3}$ in PMNS) 
associated with the bending (curvatures).
\cite{features,cornerel}.
\end{itemize}
Working systematically through the idea suggested by these examples, one 
obtained a set of criteria for calculating the mass and mixing patterns 
of both quarks and leptons once given the rotation trajectory of $\balpha$ 
and the coefficients $m_T$ (meaning the empirical mass of the heaviest 
generation) for each fermion species.   And one did get in R2M2 days quite 
sensible answers by choosing an appropriate trajectory for $\balpha$.

What the FSM now does is to give a set of differential equations for 
$\balpha$.  ($\balpha$ being a global quantity does not by itself get 
renormalized by framon loops but the vacuum does, and since $\balpha$ is 
coupled to the vacuum, as seen in (\ref{Phivac}) above, the result follows.)  
These equations depend on the coupling strength of the framon and a fudge 
parameter representing some secondary effects one could not then calculate.  
The solution of these equations yields 3 integration constants.  Applying 
the above R2M2 criteria requires $m_t, m_b, m_\tau, m_{\nu_3}$, the first 
3 given by experiment, but the Dirac mass $m_{\nu_3}$ of the heaviest 
neutrino is unknown and has to count as another parameter.  There is one 
other parameter, the strong CP angle $\theta_I$, that we shall come to in 
the next section, bringing the total to 7 real parameters.  Once given 
these 7 parameters, then the mass spectrum of all quarks and charged leptons 
together with the CKM and PMNS matrices can be calculated.  (The physical 
masses of the neutrinos depend on the see-saw mechanism but can be studied 
as well.)  One can thus make a fit to the experimental data on these 
quantities by adjusting the 7 parameters.  This was done in \cite{tfsm} and 
the result is shown in Table \ref{tfsmfit}.

\begin{table}
\centering
\begin{tabular}{|l|l|l|l|l|}
\hline
& Expt (June 2014) & FSM Calc & Agree to & Control Calc\\
\hline
&&&& \\
{\sl INPUT} &&&&\\
$m_c$ & $1.275 \pm 0.025$ GeV & $1.275$ GeV & $< 1 \sigma$&$1.2755$ GeV\\
$m_\mu$ & $0.10566$ GeV & $0.1054$ GeV & $0.2 \%$ & $0.1056$ GeV\\
$m_e$ & $0.511$ MeV &$0.513$ MeV & $0.4 \%$ &$0.518$ MeV\\
$|V_{us}|$ & $0.22534 \pm  0.00065$ & $0.22493$ & $< 1 \sigma$ &$0.22468$\\
$|V_{ub}|$ & $0.00351^{+0.00015}_{-0.00014}$& $0.00346$ & $< 1 \sigma$&$0.00346$ \\
$\sin^2 2\theta_{13}$ & $0.095 \pm 0.010$ & $0.101$ &$< 1 \sigma$ &$0.102$\\
\hline
&&&& \\
{\sl OUTPUT} &&&&\\
$m_s$ & $0.095 \pm 0.005$ GeV & $0.169$ GeV & QCD &$0.170$ GeV \\
& (at 2 GeV) &(at $m_s$) &running& \\
$m_u/m_d$ & $0.38$---$0.58$ & $0.56$ &  $< 1 \sigma$&$0.56$ \\
$|V_{ud}|$ &$0.97427 \pm 0.00015$ & $0.97437$ & $< 1 \sigma$&$0.97443$ \\
$|V_{cs}|$ &$0.97344\pm0.00016$ & $0.97350$ & $< 1 \sigma$&$0.97356$ \\
$|V_{tb}|$ &$0.999146^{+0.000021}_{-0.000046}$ & $0.99907$ &$1.65
\sigma$&$0.999075$ \\
$|V_{cd}|$ &$0.22520 \pm 0.00065$ & $0.22462$ & $< 1 \sigma$ &$0.22437$\\
$|V_{cb}|$ & $0.0412^{+0.0011}_{-0.0005}$ & $0.0429$ & $1.55 \sigma$&
$0.0429$ \\
$|V_{ts}|$ & $0.0404^{+0.0011}_{-0.0004}$ & $0.0413$ &$< 1 \sigma$& 
$0.0412$\\  
$|V_{td}|$ & $0.00867^{+0.00029}_{-0.00031}$ & $0.01223$ & 41 \% & $0.01221$\\
$|J|$ & $\left(2.96^{+0,20}_{-0.16} \right) \times 10^{-5}$ & $2.35
\times 10^{-5}$ & 20 \% &$2.34\times 10^{-5}$ \\
$\sin^2 2\theta_{12}$ & $0.857 \pm 0.024$ & $0.841$ &  $< 1 \sigma$& $0.840$\\ 
$\sin^2 2\theta_{23}$ & $>0.95$ & $0.89$ & $> 6 \%$ &$0.89$\\
\hline 
\end{tabular}
\caption{Calculated fermion masses and mixing parameters compared with
experiment taken from \cite{tfsm}} 
\label{tfsmfit}
\end{table}

The fitted trajectory for $\balpha$ is shown in Figure \ref{florosphere}, 
giving the curve traced out by $\balpha$ on the unit sphere, and in Figure 
\ref{thetaplot} giving the speed with respect to the scale $\mu$ at which 
this curve is traced.  Many of the features in Table \ref{tfsmfit} can 
qualitatively be gathered already from these figures.  For instance, one 
notices that the rotation starts off slowly at high scale (because there 
is a rotational fixed point at $\mu = \infty$) and goes faster as the scale 
decreases.  From this it follows that mass leakages and mixings, both 
being consequences of rotation, are smaller at higher scales than at low 
scales.  Hence:
\begin{itemize}
\item {\bf [qf4]} 
$m_c/m_t < m_s/m_b < m_\mu/m_\tau$, $m_u/m_t < m_d/m_b < m_e/m_\tau$,
\item {\bf [qf5]}
$V_{ts}, V_{cb} < U_{23}$, $V_{td}, V_{ub} < U_{13}$.
\end{itemize}
In other words, together with {\bf [qf1] - [qf2]} above, these cover 
already the main qualitative features of the mass and mixing patterns of 
quarks and leptons without any calculations performed.  A fit is needed 
only to check that numbers approximating those seen in experiment can 
actually be obtained.  One can claim therefore, we think, that the rest 
of the question {\bf [Q2]} posed above in Section 3 is now also answered, 
namely that
\begin{itemize}
\item {\bf [R2$'$]} the 3 generations of quarks and leptons would manifest 
the characteristic mass and mixing patterns seen in experiment.
\end{itemize}
However, that the fit should end up with such a quality, with many items
being inside the stringent experimental bounds, was a bit of a surprise, 
obtained as it was with at best only a 1-loop approximation.  In effect, 
the FSM has replaced by its 7 parameters in this fit 17 of the parameters 
in the SM, although only 12 of the latter lot have so far been measured 
in experiment and figured in Table \ref{tfsmfit}.

\begin{figure}
\centering
\includegraphics[height=17cm]{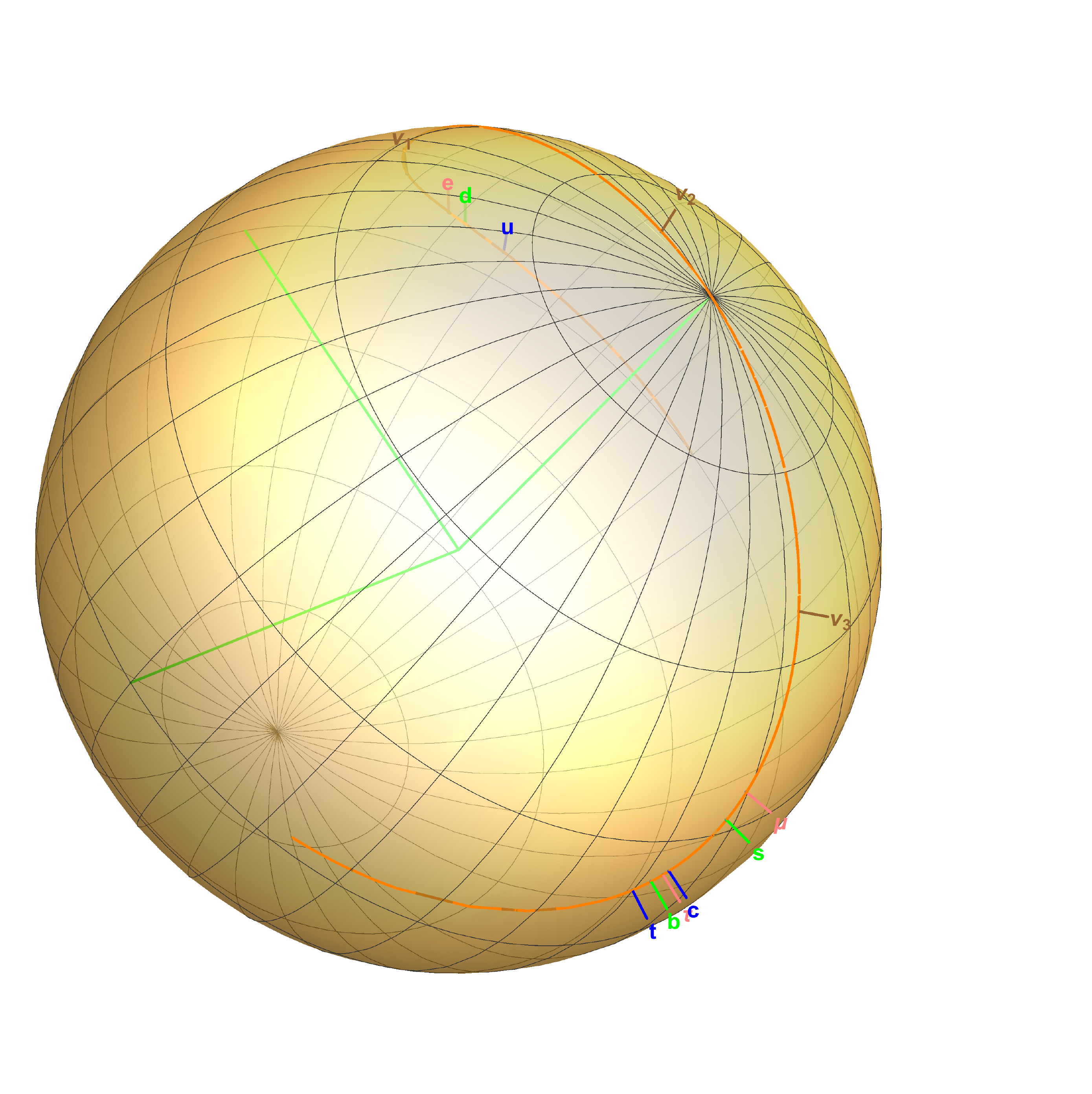}
\caption{The trajectory for $\balpha$ on the unit sphere in
generation space obtained from the parameter values given
in \cite{tfsm}, showing the locations on the trajectory
where the various quarks and leptons are placed: high scales in front.}
\label{florosphere}
\end{figure}

\begin{figure}
\centering
\includegraphics[height=7.5cm]{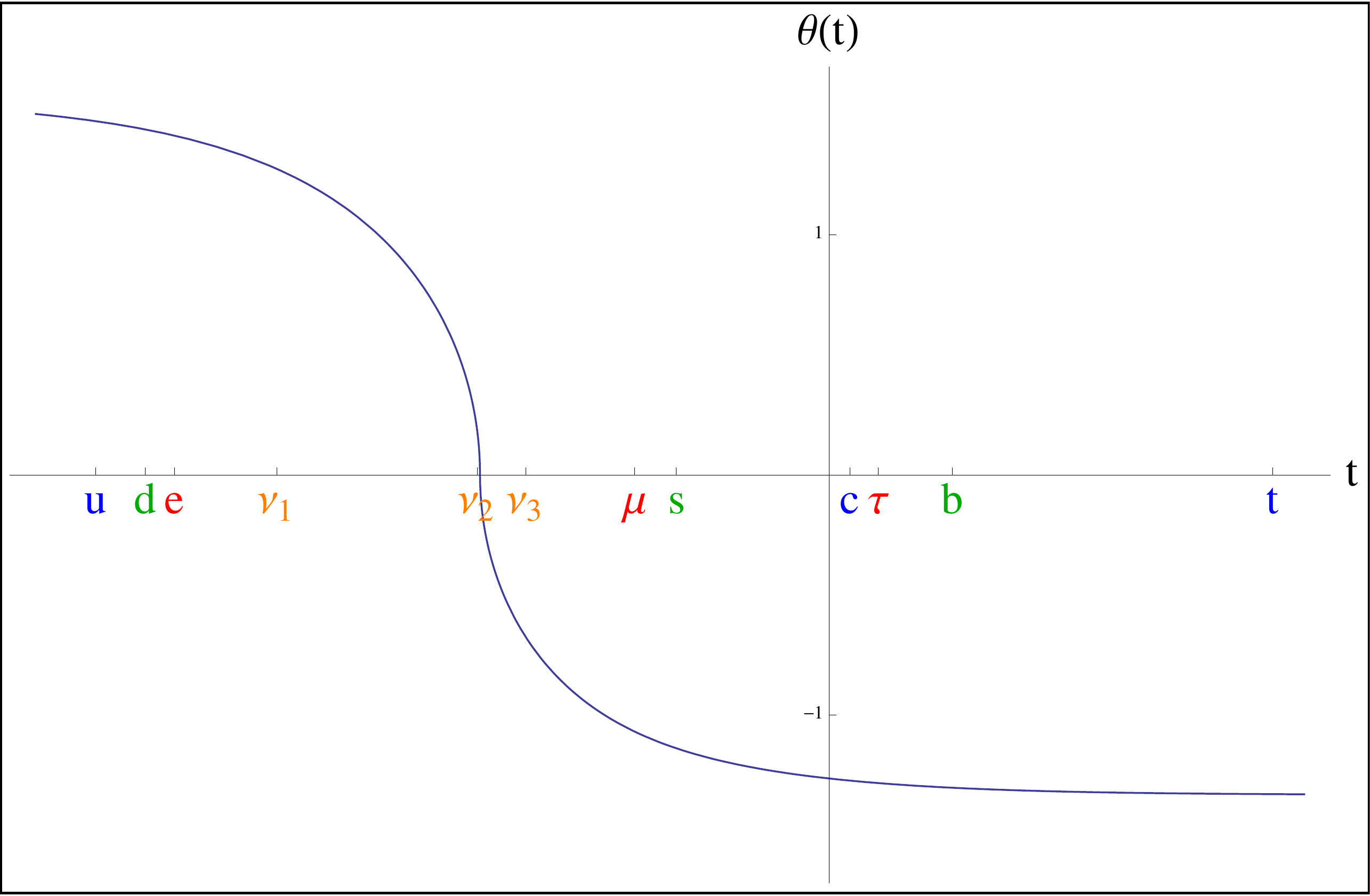}
\caption{Solution of the RGE for
$\theta$  as a function of $t= \log \mu^2$, where $\mu$ is the scale
in GeV,
obtained with parameters given in Table \ref{tfsmfit}.}
\label{thetaplot}
\end{figure}

Two points of detail in Table \ref{tfsmfit} are worth a special mention:
\begin{itemize}
\item {\bf [D1]} That a value for the Jarlskog invariant (or CP-violating 
phase) is obtained from a fit which is real as presented above is due to 
an important additional result of the FSM which will be described in the 
next section in connection with the 7th parameter $\theta_I$ mentioned a 
few paragraphs above. 
\item {\bf [D2]} That $m_u < m_d$ is, of course, a crucial empirical fact, 
without which the proton would be unstable and we, together with the world 
we know, would not exist.  The question why this should be the case has 
long been asked (even in the days before quarks were known, in the form 
why the proton is lighter than the neutron) but has never been given a 
generally accepted answer.  The mystery deepens further when the heavier 
quarks were discovered since the up versions of these are both heavier 
than the down versions: $m_t >> m_b, m_c > m_s$.  That the FSM in Table 
\ref{tfsmfit} should come up with the right result $m_u < m_d$ is thus a 
pleasant surprise.  As explained in \cite{tfsm}, this comes about in the 
FSM because of a transition point in the trajectory of $\balpha$ at which 
the geodesic curvature changes sign, a transition point which, as we shall 
see, has some further significant roles to play.
\end{itemize}

\section{Further consequence: unified approach to CP physics }

What is more, the FSM solution claimed above of the generation puzzle 
for fermions has brought with it a very significant bonus in the form 
of what can potentially become a unified treatment of all CP physics
\cite{cpslept,cpslash}.

We note first that to the SM, CP physics is in a sense an enigma for 
the following reasons:
\begin{itemize}
\item {\bf [a]}  The so-called strong CP problem that SM has simply 
ignored.
\end{itemize}
In the Lagrangian density of the QCD action, invariance considerations 
admit in principle a term, say from instantons, of the form:  
\begin{equation}
- \frac{(\theta_I)}{16 \pi^2} \tr_C (H^{\mu \nu} H^*_{\mu \nu}),
\label{thetatermC}
\end{equation}
where $H^{\mu \nu}$ is the colour gauge field, $H^*_{\mu \nu} = - \half 
\epsilon_{\mu \nu \rho \sigma} H^{\rho \sigma}$, $\tr_C$ is a trace over 
colour indices, and $\theta_I$ can take any value $0 \rightarrow 2 \pi$.
This term can lead to CP violations of order unity in strong interaction 
physics, in contradiction to experiment, for example, to the measured 
bound on the electric dipole moment (edm) of the neutron \cite{edmbound} 
which wants $\theta_I < 10^{-9}$.  To avoid this difficulty, the usual 
formulation of the SM has simply left it out.
\begin{itemize} 
\item {\bf [b]}  The quark mixing CKM matrix admits a CP-violating phase 
$\delta_{CP}$ but the SM gives no indication of its physical origin or 
its size.
\end{itemize}

Next, a flavour equivalent to (\ref{thetatermC}) of the form:
\begin{equation}
- \frac{(\theta'_I)}{16 \pi^2} \tr_F (G^{\mu \nu} G^*_{\mu \nu}),
\label{thetatermF}
\end{equation} 
where $G^{\mu \nu}$ denotes the flavour gauge field and $\tr_F$ is the 
trace over flavour indices, can in 
principle appear in the flavour theory from instantons and give 
similarly large CP-violations to leptons.\footnote{This statement may be 
controversal for some authors think that this term gives no physical 
effect.  We disagree, see \cite{cpslept} Appendix.} 
\begin{itemize}
\item {\bf [c]}  In the SM, this theta-angle term from flavour instantons  
is likewise ignored.
\item {\bf [d]}  The lepton mixing PMNS matrix admits also a CP-violating 
phase $\delta'_{CP}$ but again the SM gives no indication of its physical 
origin or its size.
\end{itemize}
Besides, in the SM no relation is known or even hinted between any of the 
4 items {\bf [a], [b], [c], [d]}.

The FSM on the other hand offers an entirely different and more coherent 
view, predicated on the form of the fermion mass matrix (\ref{mfact}) on 
which, we recall, the mass and mixing patterns also depend.  The crucial 
point is that (\ref{mfact}) is of rank 1, and has therefore at any scale 
two zero modes in directions in generation space orthogonal to $\balpha$.

Now it is well known that once there is a zero mode for quarks, say, then 
a chiral transformation on that zero mode:
\begin{equation}
\psi_0 \rightarrow \exp(-\tfrac{i}{2} \theta_I \gamma_5) \psi_0
\label{chiraltrans}
\end{equation} 
will generate in the measure of Feynman path integrals an exponential 
factor \cite{Fugikawa,Weinbergbk} with exponent of exactly the same form 
but opposite in sign to (\ref{thetatermC}) so as to cancel that term and 
hence ``solve'' the strong CP problem.  The same chiral transformation on 
a massive quark mode will of course have the same effect but it will also 
make the mass term complex and therefore CP-violating and unacceptable.

Physically, what this means is that for a massive mode, the mass 
term fixes the relative phases of the CP-conjugate pairs, while for a 
zero mode, no such constraint is there so that one is free to define CP 
differently with a chiral transformation as per (\ref{chiraltrans}) to 
keep the theory as CP-invariant as possible.  
In other words, for a theory with quark 
zero modes, CP is yet undefined up to a phase, and one is free to choose 
for convenience a phase convention in which the theory is CP-invariant,
at least at the action level. 
Hence, for the FSM, because of the zero mode inherent in the fermion mass 
matrix (\ref{mfact}), CP is intrinsically undefined, and an appropriate 
choice of the relative phase between CP-conjugate pairs of this zero 
mode will make the theory explicitly CP-invariant at the action level,
and avoid the appearance of any theta-angle term in the action.  In this 
language, such a term appears in the usual formulation only because of 
an inappropriate arbitrary choice of the said relative phase.  

Now, in the FSM with mass matrix (\ref{mfact}), there are 2 zero modes at 
any scale in the two directions orthogonal to $\balpha$ which we can take 
as $\btau$, the tangent to the trajectory of $\balpha$, and $\bnu$, the 
binormal orthogonal to both $\balpha$ and $\btau$.  It however turns out 
that a chiral transformation performed on the mode in direction $\btau$ 
would make $\balpha$ complex when it starts to rotate whereas the RGE
derived in the FSM \cite{tfsm} would want to keep it real.  But a chiral 
transformation (\ref{chiraltrans}) performed on the zero mode in the 
direction of the binormal $\bnu$ at every scale would cancel the term 
(\ref{thetatermC}) \cite{cpslept} and still keep $\balpha$ real.  Hence 
for the FSM:
\begin{itemize}
\item {\bf [a$'$]}  The strong CP problem is solved at every scale by an 
appropriate chiral transformation on the quark zero mode of (\ref{mfact})
in the direction of the binormal $\bnu$ to the trajectory of $\balpha$.
\end{itemize}
 
But, $\bnu$ being orthogonal to $\balpha$ which rotates, will rotate with 
scale as well.  This means that the chiral transformation needed to solve 
the strong CP problem will have to be performed in different directions at 
different scales.  Hence, the $c$ quark with state vector $\bc$, defined 
at the scale $\mu = m_c$ and has a component in the direction of $\bnu$ 
at that scale, will acquire a different phase from the operation in 
{\bf [a$'$]} from the phase acquired by the state vector $\bs$ of the $s$ 
quark which is defined at the different scale $\mu = m_s$ and has also 
a component in the direction of $\bnu$ at that different scale.  Thus, 
the operation {\bf [a$'$]} to solve the strong CP problem at every scale 
will make the CKM matrix element $V_{cs} = \bc^\dagger \bs$ complex.  
A similar conclusion applies to most of the other CKM elements leading 
then to the result that in the FSM \cite{atof2cps}:
\begin{itemize}
\item {\bf [b$'$]}  A CP-violating (KM) phase appears in the CKM matrix 
as a consequence of {\bf [a$'$]} because of mass matrix rotation.
\end{itemize}        
It even follows that, for a $\theta_I$ of order unity as it is expected 
to be, 
the Jarlskog invariant $J$ \cite{Jarlskog} measuring the CP-violating 
effects in the CKM matrix will automatically be of the right order 
$10^{-5}$ as observed in experiment \cite{atof2cps} given the absolute 
values of the CKM matrix elements themselves.  Indeed, in the fit to 
experiment \cite{tfsm} summarized in Table \ref{tfsmfit}, {\bf [a$'$]} 
and {\bf [b$'$]} have already been taken into account, with $\theta_I$ 
taken as the 7th parameter mentioned above in {\bf [D1]} but not then 
specified.  It is seen in Table \ref{tfsmfit} that a value for $J$ 
rather close to experiment is indeed obtained for a value of $\theta_I$ 
of order unity as claimed.
  
Essentially the same considerations can be parallelled in the flavour 
theory except that in the FSM there is an additional theta-angle term 
to (\ref{thetatermF}) coming from the operation {\bf [a$'$]}.  Quarks 
carry flavour as well as colour, so that the chiral transformation
(\ref{chiraltrans}) performed to cancel the $\theta_I$ term in colour
will generate from the measure of Feynman integrals a theta-angle term 
in flavour as well, giving in total for flavour the theta-angle term 
\cite{cpslept}:
\begin{equation}
- \frac{(\theta'_I + \theta'_C)}{16 \pi^2} \tr_F (G^{\mu \nu} G^*_{\mu \nu}),
\label{thetatermF'}
\end{equation}  
where $\theta'_C = - \tfrac{3}{2} \theta_I$.  Neverheless, following 
the same reasoning as in the colour case, one obtains:
\begin{itemize}
\item {\bf [c$'$]} Any CP-violating theta-angle term (\ref{thetatermF'}) 
can be cancelled by an appropriate chiral transformation on the lepton 
zero mode of (\ref{mfact}) in the direction of $\bnu$.
\item {\bf [d$'$]}  A CP-violating (KM-like) phase will appear in the 
PMNS matrix as a consequence of {\bf [c$'$]} because of mass matrix 
rotation.
\end{itemize}
And again it follows that for $\theta'_I$ and $\theta'_C$ of order unity, 
the Jarlskog invariant $J'$ for leptons will automatically be of the 
right order $10^{-2}$ as observed in experiment given the absolute values 
of the PMNS matrix elements.

Thus, to summarize, the FSM seems to have offered us the following new 
unified picture of CP physics:
\begin{itemize}
\item The theta-angle term and the CP-violating phase in the mixing 
matrix, whether in the colour or flavour sector, are but two facets of 
the same physics, being related just by mass matrix rotation, so that in 
colour (flavour) $\theta_I$ ($\theta_I'$) and $\delta_{CP}$ ($\delta'_{CP}$)
count as just one parameter deducible in principle from one another.
\item Quarks and leptons are treated similarly as far as CP physics is 
concerned, apart from the difference that quarks carry both colour and 
flavour while leptons carry only flavour.
\item That the Jarlskog invariants which measure the size of CP-violation 
in quarks and leptons both turn out to have the order of magnitude seen 
in experiment suggest that all known CP-physics may already be accounted 
for in the above.
\end{itemize}
This seems a considerably neater and more unified treatment of CP physics 
than that in the SM, as outlined in {\bf [a]} - {\bf [d]}.

\begin{figure}[h]
\centering
\includegraphics[scale=0.325]{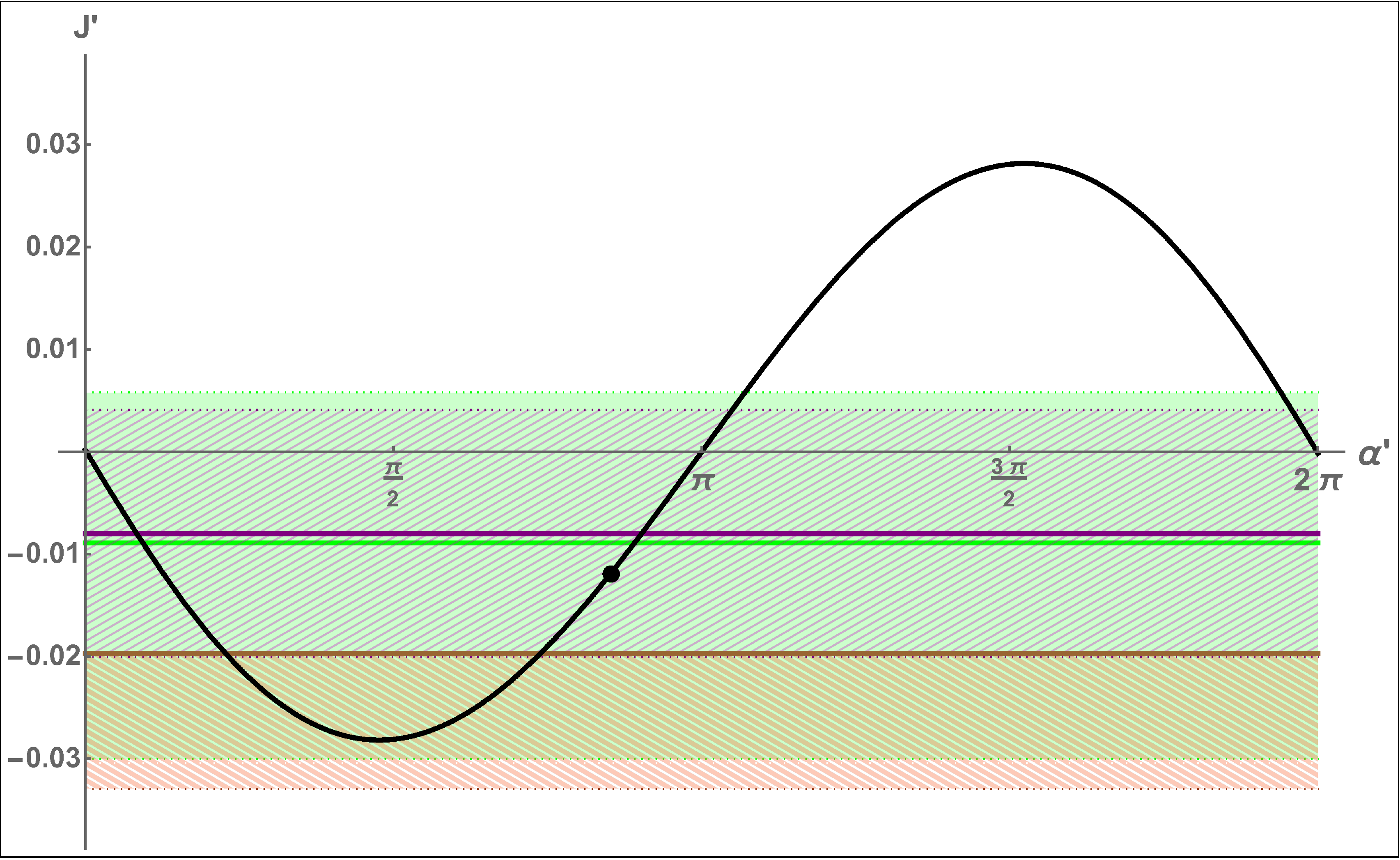}
\caption{Comparing FSM estimates to experiment.  Curve shows the
predicted values for $J'$ for various values of $\alpha'$ about 
the preferred value (dot) suggested by \cite{cpslash}.  (Here 
$\alpha'$ is the angle by which the lepton zero mode in direction 
$\bnu$ is chirally transformed.)  One notes that $J'$ is of the 
right order $\sim 10^{-2}$ for all values of $\alpha'$ except, of 
course, near $n \pi$, while for the preferred value (dot) $J'$ is 
in the range favoured by present experiment.  The lines represent 
best fit value of $J'$ given by respectively \cite{pdg} (brown), 
\cite{nufit} (green), and \cite{Valle} (mauve), while the shaded 
areas are the 1$\sigma$ range of $J'$ inferred from \cite{pdglive} 
(orange hatch), \cite{nufit} (green), and \cite{Valle} (mauve 
hatch).}  
\label{varialpha'}
\end{figure}

Two observations made in \cite{cpslash} applying such ideas on CP in the 
FSM may be worth mentioning, although they are at present only tentative:
\begin{itemize}

\item {\bf [O1]} In the predicted hidden sector some of the $F$ particles 
may undergo strong interactions similar in nature to those seen for quarks 
in our standard sector and there would be a parallel to the strong CP problem 
as well.  If one were to postulate that strong interactions should also be 
CP-invariant in the hidden as they are in our standard sector, then the 
theta-angle term (\ref{thetatermF}) from instantons with coefficient 
$\theta'_I$ should be cancelled by a chiral transformation on an $F$ zero 
mode, leaving only the $\theta'_C$ term in (\ref{thetatermF'}) to be 
cancelled by a chiral transformation on the lepton zero mode.  But this 
$\theta'_C = \tfrac{3}{2} \theta_I$ has already been given a value in the 
fit to data cited in Table \ref{tfsmfit}.  This means that the FSM would 
now be able to calculate the Jarlskog invariant $J'$ for leptons, or 
equivalently the CP-volating phase in the PMNS matrix, which is being 
measured in experiment.  An estimate, though only crude at present, made 
in \cite{cpslash} gives the following: 
\begin{equation}
J' \sim - 0.012, \ \ \  \delta'_{CP} \sim 0.011 \pi,
\label{J'}
\end{equation}
which is seen in Figure \ref{varialpha'} to sit within the range favoured by 
present experiment.

\item {\bf [O2]} One salient feature of the above treatment of CP in the
FSM is that at every scale, the theory can be made CP-invariant, but when 
the scale changes, then CP-violation can develop.  Now, in baryogenesis 
in the early universe, CP-violation plays a crucial role \cite{Sakharov},
and it is often assumed that CP is primordially violated.  However, 
if one now accepts the scenario suggested above by the FSM, CP-violation 
can develop as the scale (temperature) changes even if the universe was 
primordially CP-conserving.

\end{itemize}

\section{Modified Weinberg mixing - probing the hidden sector (I)}

Emboldened somewhat by the seemingly positive results obtained so far 
in the standard sector, let us now gather up our courage and approach 
to probe a little the hidden sector, the great unknown, starting with 
a modification to the standard Weinberg mixing of vector bosons.  In
a few email exchanges with Weinberg around a year before his decease 
about the FSM and his work, he remarked that we were ``courageous'' to 
consider changes to the standard (his) mixing scheme.   The adjective 
``courageous'' he used, we fully realized, was just a polite substitute 
for ``foolhardy'', for the standard mixing scheme for vector bosons has 
been subjected to so many intensive checks by experiment that any change 
would find it hard to survive their scrutiny.  For our part, however, 
it was neither courage nor foolhardiness but simply lack of foresight, 
as we told Weinberg.  We did not realize when we constructed the FSM 
that it would involve changes in the standard mixing, focussed, as we 
were, on the generation problem for fermions.  When we did realize the 
fact later we, of course, stopped everything to check first whether 
these changes would contradict experiment blatantly, or otherwise it 
would not be worthwhile pushing on without modifications to the model.

Let us begin by retracing the steps which led to the conclusion that 
framing colour in the FSM would modify the Weinberg 
mixing, which are in fact very similar to those by which the framing 
of flavour in the SM are seen to give the Weinberg mixing in the first 
place.  We remark first that in Section 6 when the framon {\bf [CF]} 
was defined, we had not given any justification for the values chosen 
for the electric charges $q$ of the framon field, without which the 
specification of its representation of $G$ would be incomplete.  One 
might be tempted to choose as the simplest the assignment $q = 0$ to 
all components of $\bPhi$, but this will not do, for the colour framon 
will combine with fundamental fermion fields via colour confinement to 
form the $F$s in Section 9, which will then carry fractional charges 
as the fundamental fermions.  Although the $F$s are supposed to be in 
the hidden sector and hard to access, the existence of such fractional 
charges would not have escaped detection.  Besides, which in the end 
amounts to the same thing, the gauge group (as distinguished from the 
gauge algebra) of the theory is what we called in \cite{cfsm} $U(1,2,3)$, 
a quotient of the product $U(1) \times SU(2) \times SU(3)$ by $\bbz_6$, 
which allows as representations only colour triplets, such as our framon, 
carrying $U(1)$ charges $- \third +n$, with $n$ an integer.  However, the 
dual colour $\widetilde{su}(3)$ symmetry being broken by the vacuum in 
(\ref{Phivac}) in Section 8, there is no need for all the dual colour 
components of the colour framon to have the same charge $q$.  

Whatever non-zero assignments of $q$ to the 3 dual colour components, 
when substituted into the FSM action, would lead to mixing between the 
$u(1)$ and the colour sector by virtue of the mass matrix deduced for 
the vector bosons.  A condition that one would insist on, however, is 
that the photon so obtained from this mixing should remain massless, or 
else the resultant theory would be unphysical.  A choice to satisfy this 
condition is that:
\begin{equation}
q = - \third,\ {\rm for}\ \tilde{1}, \tilde{2}; \ \ 
q = + \twothirds,\ {\rm for}\ \tilde{3},
\label{cframonqs}
\end{equation}
which was what was meant by the rather cryptic entries in (\ref{cframon}) 
in Section 6 above.  Notice that this does not claim to be the only choice 
possible to keep the photon massless, but it does seem to be the simplest, 
most analogous to the original Weinberg mixing, as can be seen in the next 
several formulae.  And so far it seems to work, as we shall see.  

Substituting the charges (\ref{cframonqs}) into the FSM action, one obtains 
the following mass submatrix between the $u(1)$ vector bosons $A_\mu$, the 
dual flavour $\widetilde{su}(2)$ vector boson $\tilde{B}^3_\mu$ and the dual colour 
$\widetilde{su}(3)$ vector boson $\tilde{C}^8_\mu$ (the rest of the mass matrix 
being diagonal in the usual labelling with Pauli and Gell-Mann matrices in 
respectively flavour and colour):
\begin{equation}
M= \left( \begin{array}{ccc}
(\ell+\third k)\,g_1^2 & 
-\ell g_1 g_2 & -\frac{k}{2\sqrt{3}}\,g_1 g_3 \\
\vspace*{1mm} \\
-\ell g_1 g_2 & \ell g_2^2 & 0\\
\vspace*{1mm} \\
 -\,\frac{k}{2\sqrt{3}} g_1 g_3 & 0 & \frac{k}{4}\,g_3^2 
\end{array} \right) 
\label{zmixedM}
\end{equation}
where
\begin{equation}
\ell = \quarter \zeta_W^2, \ \ k=\tfrac{2}{3} (1+R) \zeta_S^2
\label{zmixedM2}
\end{equation}
This has a zero mode, as expected, which we identify as the photon:
\begin{equation}
 v_1= \left( \begin{array}{c} \frac{e}{g_1} \\ \vspace*{0.3mm}\\
\frac{e}{g_2} \\\vspace*{0.3mm}\\  \frac{2}{\sqrt{3}}                          
            \frac{e}{g_3} \end{array} \right) 
\label{photonmixed}
\end{equation}
where
\begin{equation}
 \frac{1}{e^2} = \frac{1}{g_1^2} + \frac{1}{g_2^2} +
  \frac{1}{\tfrac{3}{4} g_3^2}.
\label{eg1g2g3}
\end{equation}

By solving the rest of the eigenvalue problem as is done in \cite{zmixed} 
one finds the two remaining massive eigenstates, the lower of which one 
identifies as $Z$, and the higher is a new vector boson we call $G$.  The 
whole calculation depends on the parameters $g_1, g_2, g_3, \zeta_W$ and 
$\zeta_S$, all of which except the last can be read or worked out from 
numbers given in the PDG tables \cite{pdg}.  Hence, for any assumed value  
of $\zeta_S$, the properties of $Z$ can in principle be calculated to be 
compared with experiment and checked for consistency.

There is one snag, however, namely that one cannot as yet perform in the 
FSM loop calculations in general with confidence, while the experimental 
accuracy of the relevant pieces of data up for checking is already beyond 
that achievable by calculations at the tree level.  In \cite{zmixed}, 
therefore, the following interim criterion is adopted.  
\begin{itemize}
\item {\bf [ICR]} The prediction of the FSM at tree level for any measured 
quantity is compared to the prediction of the SM for the same quantity 
(i.e. in accordance with the Weinberg mixing scheme) also at tree level.  
If the difference between these two predictions comes out to be less than 
the experimental error achieved to-date for that quantity, then we would 
consider that a pass for the FSM.
\end{itemize}  
This assumes first that the SM predictions for that quantity have already 
been checked positively against experiment at loop levels, which would be 
true in most cases.  Secondly, it assumes that the difference between the 
loop corrections of the SM (already calculated) and those of the FSM (not 
yet calculable) are of a lower order than the difference between the two 
tree-level predictions of respectively the SM and the FSM.  

If this interim criterion {\bf [ICR]} is adopted, then the comparison with 
experiment can be carried through, which is done in \cite{zmixed} with the 
following results for $\zeta_S = 2$ TeV:
\begin{eqnarray}
\Delta m_Z^{th} = 10.4\ {\rm MeV} & < & \Delta m_Z^{exp} = 15\ {\rm MeV}, \\ \nonumber
\Delta \Gamma_{e^+ e^-}^{th} = 0.028\ {\rm MeV} & < & \Delta 
\Gamma_{e^+ e^-}^{exp} = 0.12\ {\rm MeV}, \\ \nonumber
\Delta \Gamma_{hadrons}^{th} = 0.341\ {\rm MeV} & < & \Delta \Gamma_{hadrons}^{exp}
   = 2.0\ {\rm MeV},
\label{compexp}
\end{eqnarray}
where $\Delta_{\cdots}^{th}$ denote the prediction of the SM minus the prediction 
of the FSM both at tree level.  In other words, the FSM has passed the test 
according to {\bf [ICR]} both for the $Z$ mass and for $Z$ decay for $\zeta_S 
= 2\ {\rm TeV}$.  And since the $\Delta$s all decrease in value for increasing 
$\zeta_S$, this means that they will all satisfy {\bf [ICR]} for any $\zeta_S 
> 2\ {\rm TeV}$.
  
Since the differences between the SM and FSM all decrease with increasing 
$\zeta_S$, it is clear that FSM predictions will revert to those of the SM 
at large $\zeta_S$, and hence to consistency with present experiment, but 
only at the cost of their decreasing practical interest.  
That the FSM results for the $Z$ mass and $Z$ decay are so close to those 
of the SM already at $\zeta_S \sim 2\ {\rm TeV}$ as to be consistent with 
present experiment is however a little bit of a surprise, given that the 
ratio $\ell/k$ which seems a natural measure for these differences is 
$\sim 10^{-2}$ while present experimental errors are impressively of the 
order $10^{-4}$.  For the $Z$ mass, this result is traced to the fact that 
the $\ell/k$ term vanishes, while for $Z$ decay it is very small.  One has 
seen as yet no guarantee, however, that such cancellations will occur in 
the many other instances that the SM Weinberg mixing scheme has been tested 
successfully against experiment.  One can regard therefore as only tentative 
the tests (\ref{compexp}) passed so far by the FSM modified Weinberg mixing 
scheme.

\begin{figure}[ht]
\centering
\includegraphics[scale=0.2]{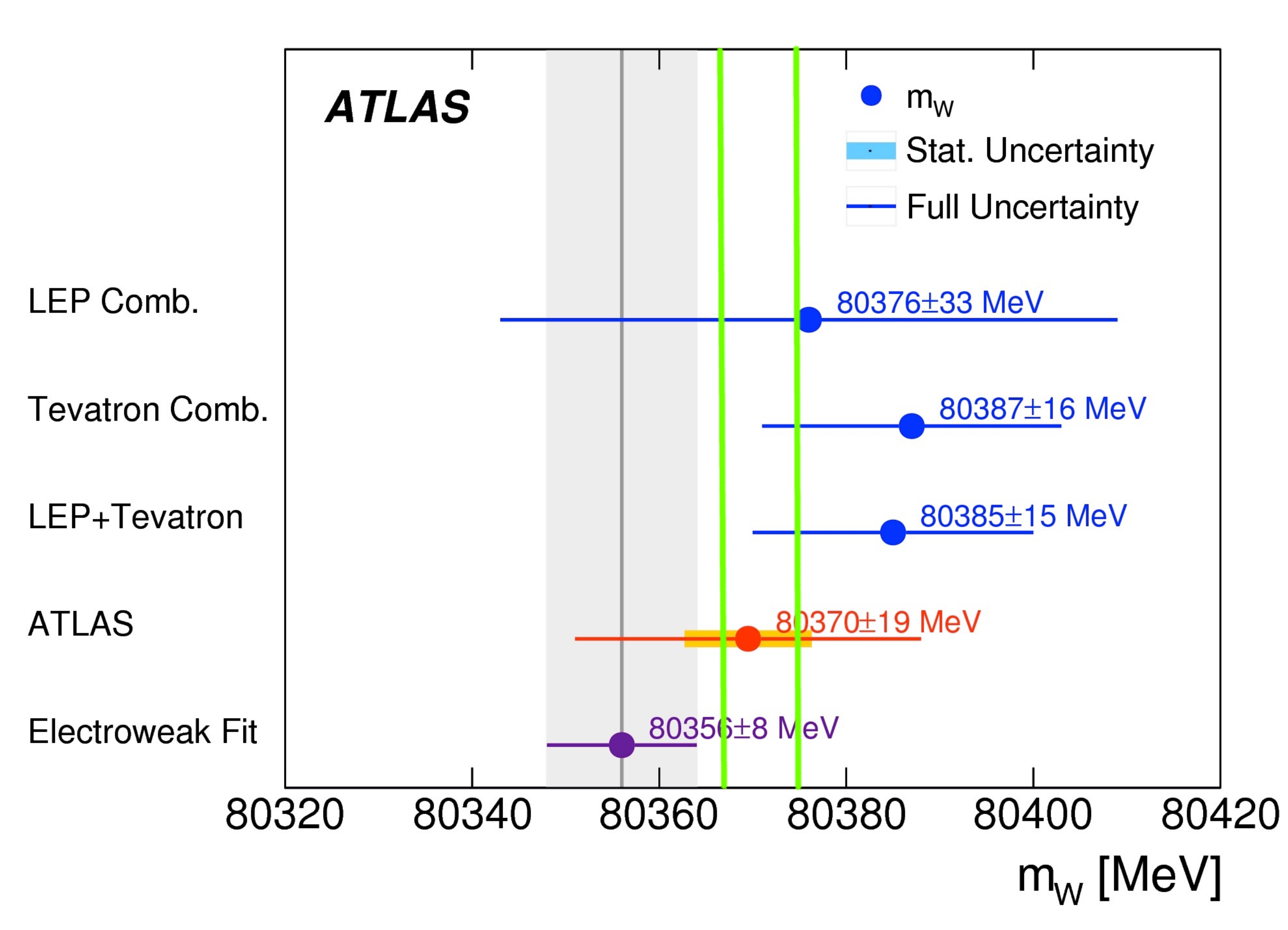}
\caption{The ATLAS measurement of the W boson mass and the combined values 
measured at the LEP and Tevatron colliders compared to the Standard Model 
prediction (mauve) and the FSM predictions (green) at $\zeta_S = 2.0$ TeV 
(left) and $\zeta_S = 1.5$ TeV (right).}
\label{mWfig}
\end{figure}

On the other hand, one can take a more positive attitude and regard these 
tests for the FSM as probes instead for the hidden sector that it predicts.  
The fits of the standard Weinberg mixing to data, though good, are not 
everywhere perfect.  Take for example again the mass difference between 
the $W$ and the $Z$.  Figure \ref{mWfig} is borrowed from \cite{Atlas}, who 
took the Weinberg mixing as predicting the $W$ mass from the $Z$ mass, 
the latter being the experimentally better measured quantity, and compared 
the prediction with experiment.  One sees there that the SM predicted value 
(i.e.\ with Weinberg mixing) is actually below what is found in experiment, 
though not statistically significantly so.  It is seen also that the FSM 
predicted value (i.e.\ with its modified Weinberg mixing) at say $\zeta_S 
\sim 2\ {\rm TeV}$ would actually fit the data better, though again not yet 
statistically significantly so.  Suppose in future when experiment is 
able to reduce further the error bars and the same situation is maintained,
then it could be regarded as a point scored by the FSM and go towards the 
support of additional mixing with a hidden sector state $G$, which would be 
a highly nontrivial result.

Of course, towards that same end, it would be better still if one can look 
experimentally for the $G$ state itself.  Given the details already known 
about the $G$, this is not a hopeless project.  However, we have so far not 
managed to gather up sufficient energy and courage to pursue it, lacking as
we do both the technical knowhow and the tools needed for the purpose.

\section{The $g - 2$ and other anomalies - probing the hidden sector (II)}

Having fortunately, at least tentatively, survived the previous test, 
let us be bolder still and probe deeper into the hidden sector.  One 
particularly striking, perhaps too daring, prediction made earlier in 
\cite{cfsm} is a bunch of particles in the hidden sector all close in 
mass around 17 MeV.  This comes about as follows.  The RGE-derived 
rotation equation for $\balpha$ which was used to obtain the fit to 
data displayed in Table \ref{tfsmfit} is coupled to another equation 
governing the scale-dependence of the ratio $R$ of (\ref{PQR}), which 
measures the relative strength of the $\widetilde{su}(3)$ 
symmetry-breaking to symmetry-restoring terms in the framon potential (\ref{V}).
Corresponding to Figures \ref{florosphere} and \ref{thetaplot} giving 
$\balpha$ as a function of the scale $\mu$, one obtains from the fit 
also Figure \ref{Rplot} giving the dependence of $R$ as a function of 
$\mu$.  One notices then that as $\theta \rightarrow 0$ at $\mu \sim$ 
17 MeV in Figure \ref{thetaplot}, $R \rightarrow 1$ in Figure \ref{Rplot}.  
Now it was found in \cite{cfsm} that there is a bunch of particles in 
the hidden sector of types $H, G$ and $F$ all with mass eigenvalues 
proportional to $\zeta_S \sqrt{1 - R}$.  Besides, it was found in 
\cite{zmixed} as quoted in the section above that $\zeta_S$ is of 
order TeV at scales of around the $Z$ mass.  This means that these 
eigenvalues $\zeta_S \sqrt{1 - R}$ must, according to Figure \ref{Rplot}, 
vanish rather abruptly as $\mu$ nears 17 MeV.  Now, physical masses of
particles are supposed to be measured at their own mass scales, namely 
as solutions to the equation:
\begin{equation}
m_X(\mu) = \mu.
\label{mphys}
\end{equation}
Given the preceeding remarks, there must be solutions to (\ref{mphys}) 
very near to 17 MeV for the particles under discussion.

\begin{figure}
\centering
\includegraphics[height=7.8cm]{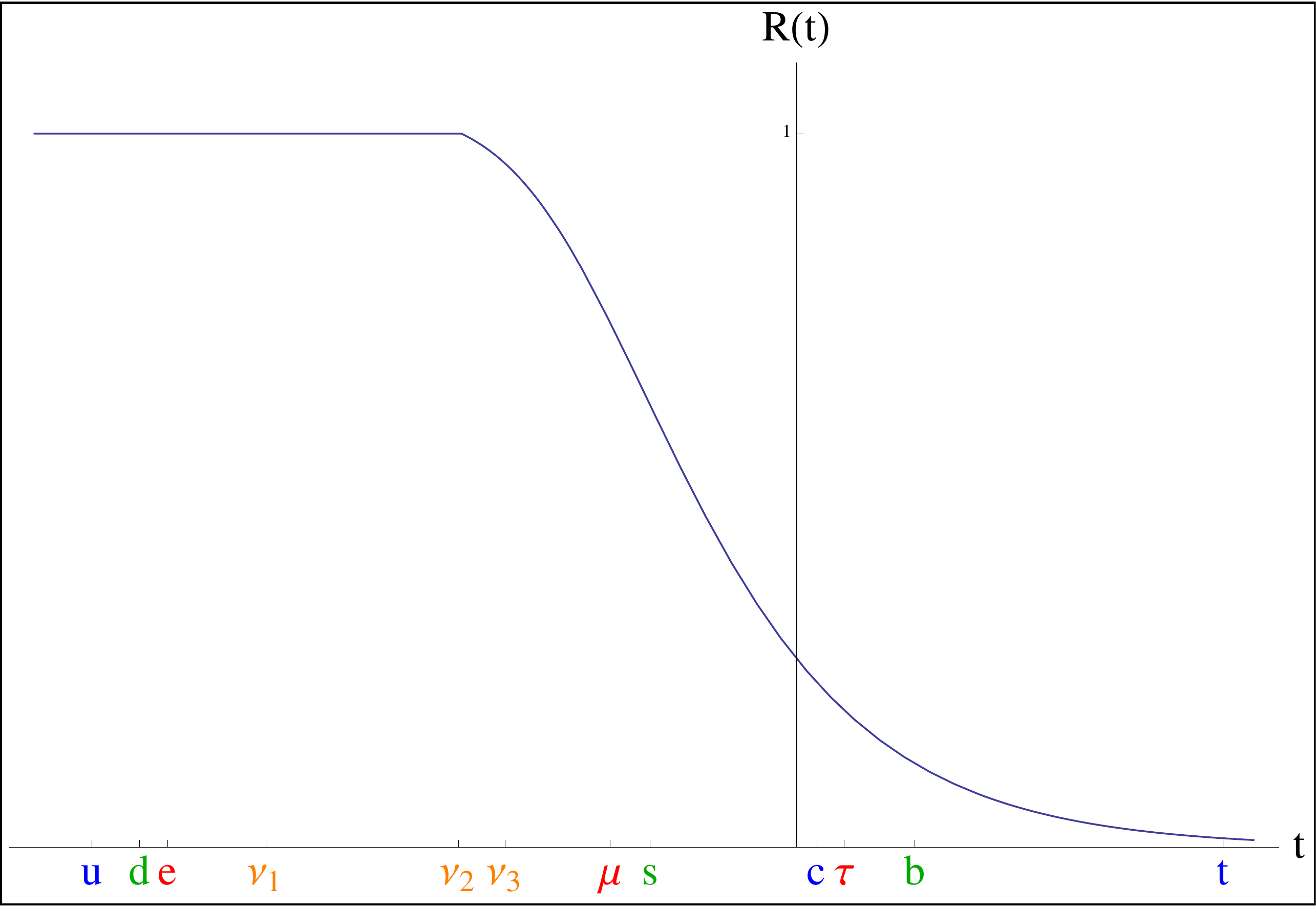}
\caption{Solution of the RGE  
for $R$  as a function of $t= \log \mu^2$, where $\mu$ is the scale in
GeV,  
obtained with parameters given as in Table \ref{tfsmfit}.}
\label{Rplot}
\end{figure}

That, based on the above arguments, there are numerous members of the 
$H, G, F$ spectrum with physical masses crowded around but just above 
17 MeV seems an audacious conclusion to draw, in a system where the 
scale characterized by $\zeta_S$ was estimated in \cite{zmixed} to be 
of order TeV.  Besides, the criterion (\ref{mphys}) for determining 
the physical mass, though commonly accepted, has, as far we we know, 
no real theoretical justification.  And though systematically employed 
in the study of the standard sector reported in Table \ref{tfsmfit} 
with apparent success, it has never been tested, as far as we know, in 
situation such as the present, when $m_X(\mu)$ has rapid dependence on 
$\mu$.  Nevertheless, given the situation adopted at present by the FSM, 
it is also a conclusion which would be logically hard to avoid, and one 
that can do with some phenomenological support.

To test the above predication of $H, G, F$ particles with masses bunched 
around 17 MeV which, if true, would be a major feature of the hidden 
sector, will not be easy phenomenologically since these particles have 
no direct communication with us residing in the standard sector.  There 
are 2 loopholes, however:
\begin{itemize}
\item A $0^+$ particle we call $H_+$ which mixes at tree level with the 
standard model Higgs boson $h_W$,
\item A $1^-$ particle we call $G_3$ which mixes with the photon $\gamma$
at 1-loop level.
\end{itemize}
These mixings will allow them to couple into the standard sector and 
manifest themselves as deviations from the SM or anomalies \cite{cfsm}.

Now, as mentioned already in Section 1, there have begun to appear in 
experiment in recent years some deviations from the SM which have 
excited a lot of interest in the community and which will also be 
relevant to us in the present context.  We note, however, a slight 
difference in attitude between those models which are specifically 
created for explaining these anomalies and the FSM which was initially 
constructed for a difference purpose, namely the generation problem 
of fermions, and only drafted later into service for understanding 
the anomalies.  Hence, the FSM, carrying with it already a fair 
amount of baggage from its earlier applications, cannot be freely 
adapted to accommodate the anomalies as some other models can.  But, 
if it nevertheless manages to accommodate the anomalies, they will 
serve for the FSM as empirical support.

\subsection{The $H_+$ state}
 
Consider now $H_+$ which, according to \cite{cfsm} mixes with the 
standard Higgs boson $h_W$ and with another $H$ state called $H_3$ 
in the tree-level mass matrix, where $H_3$ has an estimated mass of 
order TeV, and for immediate purposes can be ignored.  The remaining 
2 states, $m_{h_W} \sim 125$ GeV, $m_{H_+} \sim 17$ MeV are 
linked by the coupling $\nu_2$ from the framon potential $V$, which 
is yet unknown, though being dimensionless can be taken as of order 
unity.  The lower mixed state $U$ will thus acquire a component in 
$h_W$ which will be small, given the large difference in value 
between $m_{h_W}$ and $m_{H_+}$, and allow it to couple to standard 
particles.  The mixing will also slightly shift its mass.  However, 
given that some parameters are still unknown, and especially with 
unknown scale-dependence, it is at present not possible to go any 
further than 
\begin{itemize}
\item {\bf [UPR]} $U$ ($J^P = 0^+$) is expected to have mass around, 
say, 20 MeV and a small coupling to standard sector particles.
\end{itemize}

By virtue of the small coupling acquired via mixing with $h_W$, $U$ 
can appear in the diagrams of Figures \ref{g-2} and \ref{Lambsfig}} and 
contribute towards respectively the magnetic dipole moment of the 
muon (electron) and the Lamb shift in muonic (electronic) hydrogen.  
Now, anomalies have indeed been reported both in the magnetic dipole 
moment of the muon and in the Lamb shift in muonic hydrogen, but not 
for the electron in either case.  Further, an anomaly is found in 
the Lamb shift of muonic deuterium as well, although again not in 
ordinary (electronic) deuterium.  The so-called $g-2$ anomaly, first 
discovered in Brookhaven \cite{g-2exptmuon},
has been confirmed at about the same level last year by a new $g-2$ 
experiment at Fermilab \cite{g-2Fermilab} which is still continuing 
to take data, and appears now quite creditable as a $3$-$4 \sigma$ effect.  
As for the anomalies 
in Lamb shifts, experts are still not entirely agreed as to their 
actual significance \cite{LSmuonichydrogen1,LSmuonichydrogen2,LSmuonicdeuterium1,
LSmuonicdeuterium2}.  
We take here the existing data at their surface 
value.  The actual numbers we used can be found in \cite{fsmanom}.
 
\begin{figure}
\centering
\includegraphics[scale=0.4]{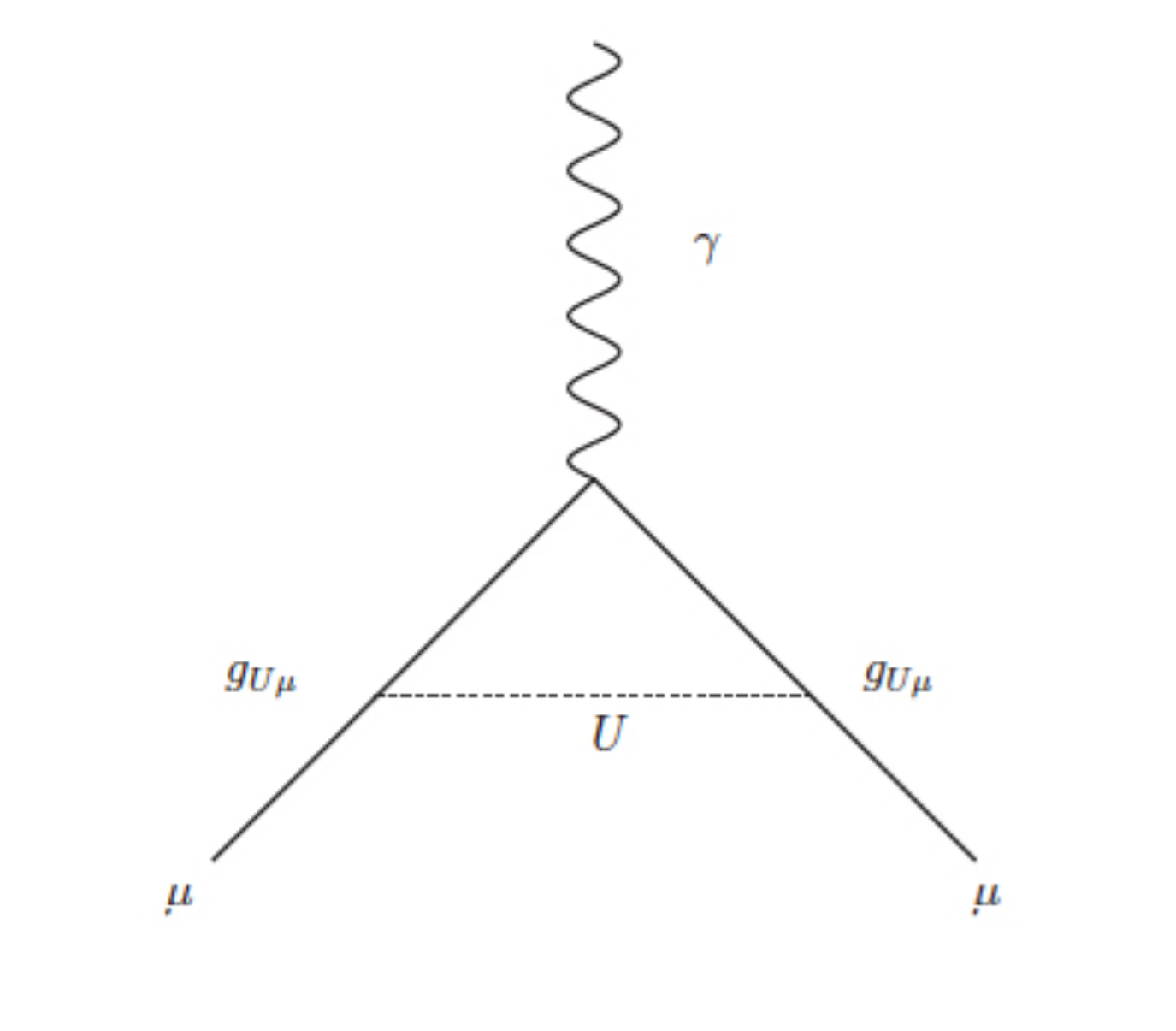}
\caption{Feynman diagram contribution of $U$ to $g-2$ anomaly of $\mu$}
\label{g-2}
\end{figure}

\begin{figure}
  \centering
 \includegraphics[scale=0.3]{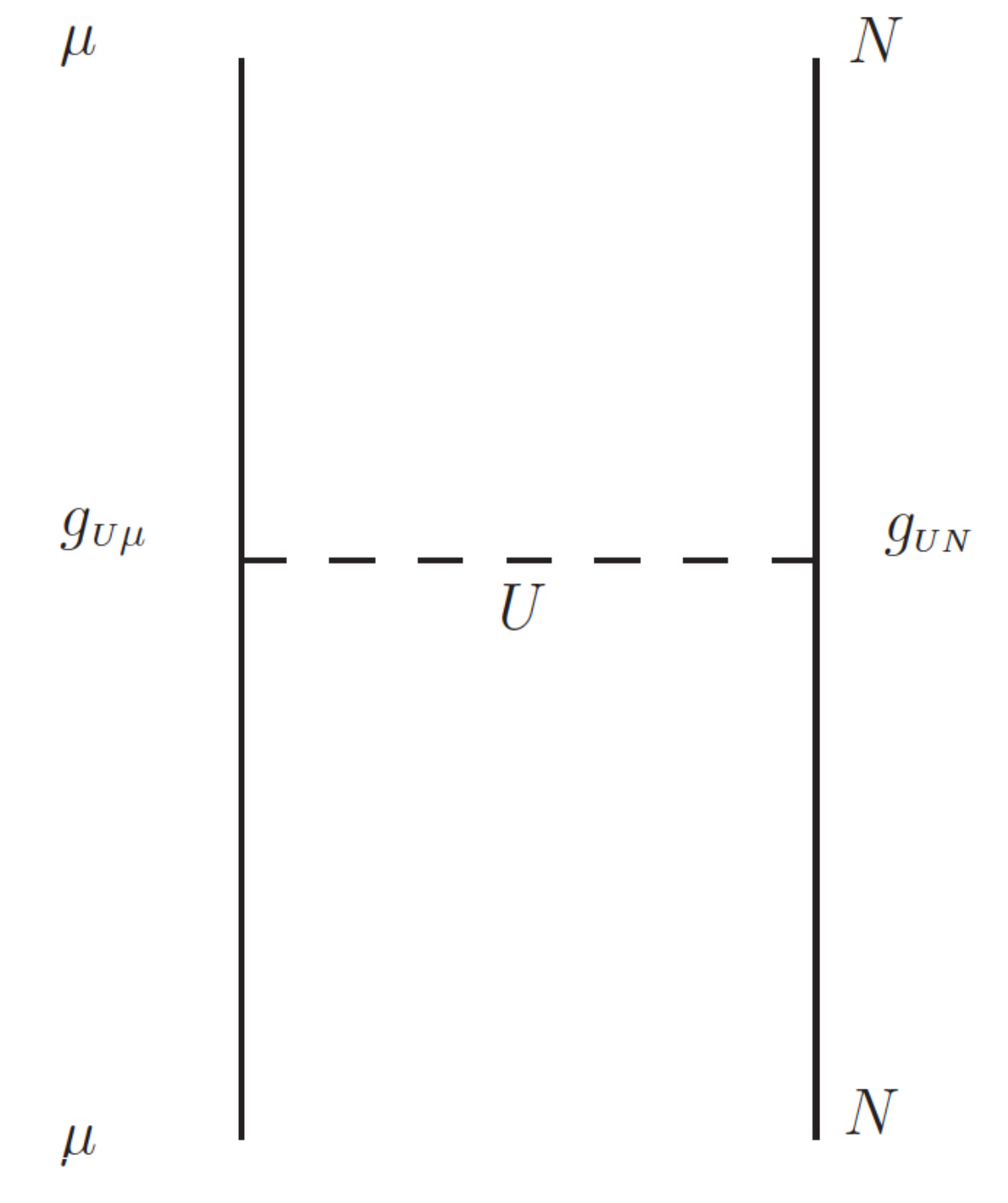}
\caption{Feynman diagram contribution of $U$ to the Lamb shift
  anomaly of muonic hydrogen and deuterium}
\label{Lambsfig}
\end{figure}

Our object is to see whether the $U$ state suggested by the FSM can 
give an overall explanation for these various bits of data, including 
(i) why the anomalies should appear in muons but apparently not in 
electrons, and (ii) why they should have the sizes they have when they 
do appear.

The answer to (i) seems mostly kinematic.  The contribution of Figure
\ref{g-2} to the $g-2$ anomaly $\Delta a_\ell = (g_\ell - 2)/2$ of 
lepton $\ell$ is explicitly (see references cited in \cite{fsmanom}): 
\begin{equation}
\Delta a_{l} \, = \, \frac{g_{U\ell}^2}{8 \pi^2} \,  \int_0^1 dz \, 
\frac{(1+z)(1-z)^2}{(1-z)^2 + z \left( \frac{m_{U}}{m_\ell} \right)^2}.
\label{Deltaal}
\end{equation} 
The contribution of Figure \ref{Lambsfig} to the splitting of the 2S-2P 
energy levels of leptonic hydrogen is given by (see references cited in
\cite{fsmanom}):
\begin{equation}
\delta E_\ell^A \, = \, -  \,g_{U\ell} \, g_{UN} \, 
\frac{m_{U}^2 a_{0N}}{8 \pi\left(1+m_{U} a_{0N}\right)^4}
\label{deltaElN}
\end{equation}
where $g_{UN}$ is the coupling of $U$ to the nucleus $N$, with mass 
$m_N$ and charge $Z$, of atom $A$ for which the Lamb shift is being 
considered, and
\begin{equation}
a_{0N} \, = \, \frac{1}{Z \alpha} \left( \frac{1}{m_N} + \frac{1}{m_\ell} 
   \right)
\label{Bohrrad}
\end{equation}
is the Bohr radius, where $\alpha$ is the fine structure constant.  
Given then the large mass ratio $m_\mu/m_e \sim 200$ between the 
muon and the electron, both the $g-2$ anomaly for the electron and 
the Lamb shift anomaly for electronic hydrogen (and deuterium) would
be substantially reduced, almost to a level not yet detectable by 
experiment.  It will be noted that there is a further suppression 
from the ratio in couplings $g_{Ue}/g_{U\mu}$.  

To actually evaluate the diagrams Figures \ref{g-2} and \ref{Lambsfig} 
so as to compare with the muon data, one will need the couplings of 
$U$ to the muon, the proton and the deuteron.  Our conclusion above
was that $U$ acquires these couplings only via its component in $h_W$ 
through mixing, in other words:
\begin{equation}
g_{Ux} = \rho_{Uh} g_{hx},
\label{gUx}
\end{equation}
where the mixing parameter $\rho_{Uh}$ is small and $g_{hx}$ is the 
coupling of $h_W$ to $x$ but taken at the scale $\mu = m_U$.  Taking 
the Yukawa coupling for quarks and leptons which gave the mass matrix 
(\ref{mfact}) at tree-level used to good effects in Section 10, and 
expanding to first order fluctuations, one obtains:
\begin{equation}
\mathcal{L}_Y \, = \, - 
\sum_{T=U,D,L,\nu} \frac{m_T}{\zeta_W} \sum_{i=1,2,3} 
|\langle \balpha(\mu) |f_i \rangle|^2 \, \, \overline{f_i}{f_i} \, h_W.
\label{hcouplings}
\end{equation}
where $m_U = m_t, m_D = m_b, m_L = m_\tau$ and $\zeta_W \sim 246$ GeV.
The state vectors $|f_i \rangle = |u \rangle, |d \rangle, |\mu \rangle$ 
relevant for the present discussion are given also in the fit of 
\cite{tfsm} summarized in Section 10, as well as $\balpha$ as a function 
of $\mu$ in Figures \ref{florosphere} and \ref{thetaplot}.  In other words, 
the couplings of $h_W$ to $u, d, \mu$ at the required scale $\mu = m_U$ 
can be read off directly.  To deduce the couplings of $h_W$ (and hence 
of $U$) to the proton and the deuteron needs nonperturbative physics,
which we took from lattice or phenomenological calculations available 
in the literature (see references in \cite{fsmanom}). 

\begin{figure}
  \centering
\includegraphics[scale=0.5]{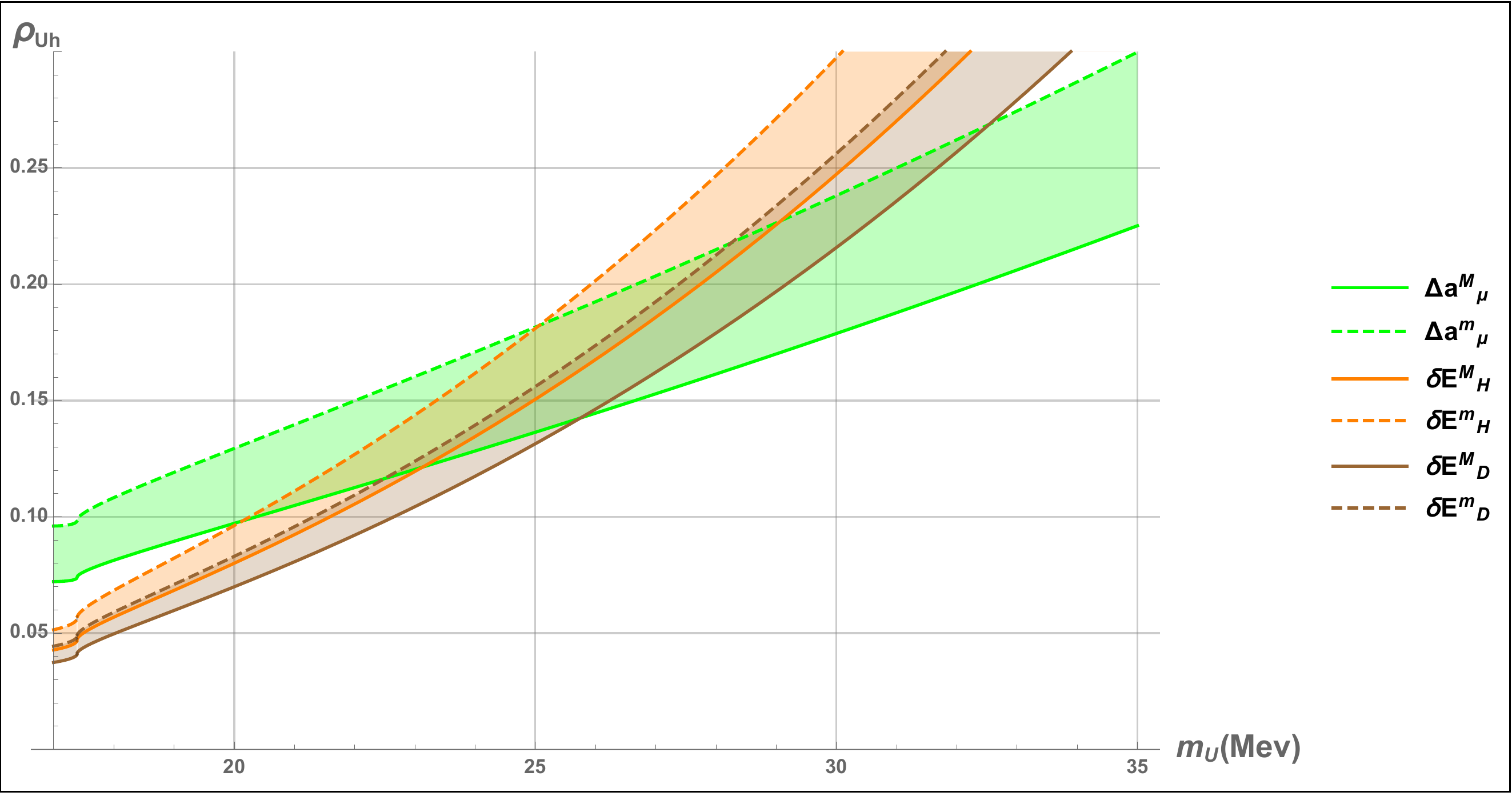}
\caption{Experimental  bounds on the mixing parameters  $\rho_{Uh}$
  from muonic $g-2$ (green), Lamb shift in muonic hydrogen (orange),
  and Lamb shift in muonic deuterium (brown), where the allowed regions
are shaded}
\label{Lambshiffc}
\end{figure}

Substituting all these into (\ref{Deltaal}) and (\ref{deltaElN}) then allows 
one to evaluate for various values of $\rho_{Uh}$ and $m_U$ respectively 
the $g-2$ anomaly in muons and the Lamb shifts in muonic hydrogen and 
deuterium.  The result is displayed in Figure \ref{Lambshiffc}.  One notes 
that the bands representing the allowed regions from the 3 separate sets 
of experiments have an overlap exacly in the range where $m_U$ is around 
20 MeV and $\rho_{Uh}$ is small (of order $10^{-1}$), both as expected in 
{\bf [UPR]}.  In other words, if one were to choose at random $m_U \sim 
20$ MeV and adjust a single parameter $\rho_{UH}$ around a sensible value 
$\sim 10^{-1}$ within the range suggested in {\bf [UPR]}, one would not be 
too far off the mark in getting both the $g-2$ and the Lamb shift anomaies 
right both for muons and electrons, and for the Lamb shift anomaly for 
both hydrogen and deuterium.  One can thus claim that the FSM can not only 
accomodate the $g-2$ and Lamb shift anaomalies as far as known at present 
but even explain to some extent their existence.

\subsection{The $G_3$ state}

We turn next to the other state $G_3$ which, as already noted, mixes 
also into the standard sector and so has a chance of being observable 
by present experiment.  It mixes, however, only at the 1-loop level 
with $G_8$ \cite{fsmanom} which is itself a component of the photon, 
as explained in \cite{cfsm}, and so the resultant mixed state, say $X$, 
of $G_3$ with $G_8$ would acquire a small component coupled via the 
photon to particles in our standard sector, in particular to electrons.
Since the mixing is 1-loop (not like the mixing of the $U$ above which 
was at tree level) and therefore small, the mass of $X$ will stay close 
to that of $G_3$, namely 17 MeV.  At that mass then, the $X$ can decay 
via the photon coupling to electron pairs with a small width but hardly 
anything else.  Again, as for the $U$, one is not yet wise enough to 
actually calculate the amount of mixing, so it is safe only to conclude:
\begin{itemize}
\item {\bf [XPR]} $X$ ($J^P = 1^-$) is expected to have mass very close 
to 17 MeV and will decay into $e^+ e^-$ with a narrow width.
\end{itemize}

It was at about this point of reasoning when writing \cite{cfsm} that 
we heard in a seminar about the Atomki anomaly discovered a short time 
before.  The $e^+ e^-$ spectrum from excited ${}^8{\rm Be}^*$ decay 
studied in \cite{Atomkiexpt} shows a $6.8\ \sigma$ bump above the 
expected background at the electron-positron invariant mass of
\begin{equation}
m_{e^+ e^-} = 16.70 \pm 0.35\ {\rm (stat)} \pm 0.5\ {\rm (sys)\ MeV},
\label{mAtomki}
\end{equation}
suggestive of a new boson $X$ being produced:
\begin{equation}
{}^8{\rm Be}^* \rightarrow {}^8{\rm Be} \  X, \ \ X \rightarrow e^+ e^-,
\label{Be*decay}
\end{equation}
with 
\begin{equation}
  \frac{\Gamma({}^8{\rm Be}^* \rightarrow {}^8{\rm Be}\ X)}{
    \Gamma ({}^8{\rm Be}^*
    \rightarrow {}^8{\rm Be}\ \gamma)}
   \br (X \rightarrow e^+ e^-) = 5.8 \times 10^{-6}.
\label{brAtomki}
\end{equation} 
This was exciting, since the mass, the decay mode and the narrow width 
are all as expected.  And, although experiment gives no clear hint as 
yet to the $J^P$ assignment of $X$, the proposed one $J^P = 1^-$ for 
$G_3$ is at least admissable, unlike the $0^+$ of $U$ of the previous 
subsection, which is not admissable if parity is to be conserved.  
True, the predicted mass 17 MeV was obtained \cite{tfsm} only in a 
numerical fit to which value no error estimate was, or even could be, 
given, so the close agreement with \cite{Atomkiexpt} could just be 
accidental.  Nevertheless, the value 17 MeV was obtained years before 
the Atomki result.

However, life is never that simple, unfortunately.  There are 2 urgent 
questions which need answers:
\begin{itemize}
\item In order to be produced in $Be$ decay, $X$ must have other ways 
of coupling to the standard sector than that gained from the mixing 
with the photon suggested above, which would be far too weak.  What are 
they, and where do they come from?
\item If such other couplings exist for $X$, then why is $X$ not seen 
somewhere else?
\end{itemize}  
These are questions that every phenomenologist dealing with $X$, 
it seems, will have to answer, and it feels at times like a game of 
hide-and-seek.  Our own answer \cite{fsmanom} is as follows.

We recall that in the FSM, $G_3$ appears as a $p$-wave bound state of 
(colour) framon-antiframon pair via colour confinement.  It interacts 
with other hidden sector particles $H, G, F$ perturbatively via the 
couplings listed in the appendices of \cite{cfsm}, of which its mixing 
with the photon is one result.  However, what will happen when a $G_3$ 
enters into a hadron which itself is a bound state of quarks via clour 
confinement?  Inside the hadronic matter, where colour is in a sense 
already deconfined, will the colour-confined state $G_3$ not sort of 
dissolve and find other ways of communicating with the other coloured 
constituents therein?  Introducing then on this intuitive basis a new 
parameter $\kappa_N$ to represent this imagined ``soft'' coupling of 
$G_3$ to nucleons, we had of course no trouble to acquire a correct 
production rate for $X$ in beryllium decay.  We could also, on the 
same basis, satisfy all the constraints \cite{fsmanom} found at that 
time in the literature set by the absence of $X$ in other experiments.
However, this last is just pure phenomenology, for at present one can 
neither justify further the advent of $\kappa$ nor estimate its value 
in FSM, $\kappa$ being in essence a nonperturbative phenomenon.

Thus, in all, in spite of the dramatic coincidence of {\bf [XPR]} with 
what is observed, one can say no more at present than just that the 
FSM can accommodate $X$, the Atomki anomaly.  Besides, although the 
$X$ has been seen again in the decay of another nucleus ${}^4{\rm He}^*$, 
or specifically in the nuclear reaction ${}^3{\rm H}(p, e^+ e^-) {}^4{\rm He}$, 
at very nearly the same mass, namely $16.94 \pm 0.12$ MeV, by the 
same group \cite{Atomki2} it needs yet to be independently confirmed. 

----------------------------------------------------------------------

The results obtained in the analysis of the last 2 subsections on $U$ 
and $X$ both in a sense serve a double purpose.  In one direction, they offer 
explanations based on the FSM for the anomalies reported in experiment,
joining thus many other suggestions in the literature for doing so .  
In the other direction, however,they provide phenomenological support 
for 2 quite audacious predictions of the FSM with some potentially very 
far-reaching consequences as noted in Section 9, namely: 
\begin{itemize}
\item first on the existence of a hidden sector, and 
\item secondly on that of a bunch of hidden sector states with masses 
near 17 MeV.
\end{itemize}
Together with the hints noted before in Section 12 and in \cite{cpslash} 
mentioned in Section 11, some inroads have now apparently been made into 
the hidden sector which is no longer as seemingly impenetrable as before.

\section{Remark}

But that, unfortunately, is as much as one has managed to dig out so 
far from the deep mine that the FSM has opened.  One feels that one 
has barely begun to scratch on the surface.  On the one hand, one has 
yet to ascertain that the FSM has not lost the wide range of physics 
that the SM has so succesfully described.  On the other, one has yet 
to find out what physics the new hidden sector contains, whether for 
example it can explain the dark matter that dominates our universe.  
It looks like a programme that would last for decades, not only for us 
but for the community, i.e.\ assuming that the FSM would survive that
long.

\section{A few more words}

We end with a few words to thank Professor Yang for giving us all such 
a rich theme to contemplate and play variations on, and to wish him a 
very happy 100th birthday with many more happy and healthy years to 
come.


\begin{thebibliography}{99}

\bibitem{ym} CN Yang and RL Mills, Phys.\ Rev.\ 95, 631 (1954).

\bibitem{g-2exptmuon} G. W. Bennett et al. [Muon $g-2$ Collaboration], Phys. Rev. D73, 
072003 (2006);  \\
doi:10.1103/PhysRevD.73.072003; arXiv:hep-ex/0602035.

\bibitem{g-2Fermilab} B Abi et al. (Muon $g-2$ Collaboration), 
  Phys. Rev. Lett. 126, 141801 (2021); arXiv:2104.03281.



\bibitem{Bdecay} LHCb Collaboration, arXiv:2103.11769.

\bibitem{tHooft} G. 't~Hooft, Acta Phys. Austr., Suppl. 22, 531  (1980).

\bibitem{Weinbergint} Interview with Steven Weinberg by the CERN Courier:
https://cerncourier.com/model-physicist/

\bibitem{Weinberggp} Steven Weinberg, Phys.\ Rev.\ D101, 035020l; 
   arXiv:2001.06582.

\bibitem{mbsm} Chan Hong-Mo and Tsou Sheung Tsun, 
  Eur. Phys. J. C52, 635-663 (2007); arXiv:hep-ph/0611364.

\bibitem{tfsm} Jos\'e Bordes, Chan Hong-Mo and Tsou Sheung Tsun,
Int. J. Mod. Phys. A30 (2015) 1550051;
doi:10.1142/S0217751X15500517; arXiv:1410.8022.

\bibitem{cfsm} Jos\'e Bordes, Chan Hong-Mo and Tsou Sheung Tsun,
Int. J. Mod. Phys. A33 (2018) 1850195;
doi:10.1142/S0217751X18501956; arXiv:1806.08268.

\bibitem{pdg} P.A. Zyla et al., (Particle Data Group), 
Prog. Theor. Exp. Phys. 2020, 083C01 (2020) and 2021 updates; http://pdg.lbl.gov/

\bibitem{Bj} James Bjoken, Ann. Phys. (Berlin) 525 (2013) A67-A79;
  DOI:10.1002/andp.2013.00724; see also the website: bjphysicsnotes.com

\bibitem{r2m2} Michael J Baker, Jose Bordes, Chan Hong-Mo and Tsou
  Sheung Tsun, Int. J. Mod. Phys. A26 (2011) 2087-2124,
  arXiv:1103.5615.

\bibitem{features} Jos\'e Bordes, Chan Hong-Mo, Jakov Pfaudler and Tsou Sheung Tsun,
  {\em Phys.\ Rev.} {\bf D58}\,(1998)\,053006; hep-ph/9802436. 

\bibitem{cornerel} Michael J Baker, Jose Bordes, Chan Hong-Mo and Tsou
  Sheung Tsun,
EPL 102 (2013) 41001, arXiv:1110.5951.

\bibitem{cpslept} Jos\'e Bordes, Chan Hong-Mo and Tsou Sheung Tsun, 
IJMP A36 (2021) 2150236; arXiv:2107.05420

\bibitem{cpslash} Jos\'e Bordes, Chan Hong-Mo and Tsou Sheung Tsun, 
IJMP A36 (2021) 2150238; arXiv:2109.11391

\bibitem{edmbound} V.\ Baluni, Phys.\ Rev.\ 19,2227 (1978); R.J.\ Crewther, P.\ Di Vecchia,
  G.\ Veneziano, and E.\ Witten, Phys.\ Lett.\ 88B, 123 (1979).

\bibitem{Fugikawa} K.\ Fujikawa, Phys.\ Rev.\ Lett.\ 42, 1195 (1979).

\bibitem{Weinbergbk} S. Weinberg, {\it The Quantum 
   Theory of Fields II} (Cambridge University Press, New York, 1996).

\bibitem{atof2cps} Jos\'e Bordes, Chan Hong-Mo and Tsou Sheung Tsun,
    Int. J. Mod. Phys.  A25 (2010) 5897-5911; arXiv:1002.3542
    [hep-ph].

\bibitem{Jarlskog} C.\ Jarlskog, Z.\ Phys.\ C 29, 491 (1985); Phys.\ Rev.\ 
  Lett.\ 55, 1039 (1985).

\bibitem{nufit} Ivan Esteban et al.,  
J. High Energ. Phys. 09 (2020) 178; arxiv.org/abs/2007.14792; www.nu-fit.org

\bibitem{Valle} P.F. de Salas et al.,  
J. High Energ. Phys. 02 (2021) 071; rxiv.org/abs/2006.1123

\bibitem{pdglive} 2021 Review of Particle Physics:
P.A. Zyla et al. (Particle Data Group), Prog. Theor. Exp. Phys. 2020, 
083C01 (2020) and 2021 update.

\bibitem{Sakharov} A.D. Sakharov, 
JETP Lett. 5. 24(1967). (Usp. Fiz. Nauk 161,61-64 (1991), English version),
doi: 10.1142/9789812815941-0013

\bibitem{zmixed} Jos\'e Bordes, Chan Hong-Mo and Tsou Sheung Tsun,
Int. J. Mod. Phys. A33 (2018), 1850190; 
doi:217751X18501907; arXiv:1806.08271.

\bibitem{Atlas} M. Aaboud et al.
(ATLAS Collaboration). Eur. Phys. J. C (2018) 78:110. arXiv:1701.07240.
Also at https://phys.org/news/2016-12-atlas-mass-lhc.html

\bibitem{LSmuonichydrogen1}
R. Pohl et al., Nature 466, 213 (2010), doi:10.1038/nature09250.

\bibitem{LSmuonichydrogen2}
A. Antognini, et al., Science 339, 417 (2013), \\doi:10.1126/science.1230016 .

\bibitem{LSmuonicdeuterium1}
Julian J. Krauth, Marc Diepold, Beatrice Franke, Aldo Antognini, Franz Kottmann, Randolf Pohl,
Annals of Physics 366 (2016) 168. \\doi:10.1016/j.aop.2015.12.006; 
arXiv:1506.01298 [physics.atom-ph].
  
\bibitem{LSmuonicdeuterium2}
C. G. Parthey et al., Phys. Rev. Lett. 104, 233001 (2010). 
\\doi:10.1103/PhysRevLett.104.233001.

\bibitem{fsmanom} Jos\'e Bordes, Chan Hong-Mo and Tsou Sheung Tsun, 
Int. J. Mod. Phys. A34 (2019), 1950140,
doi:10.1142/S0217751X19501409; arXiv:1906.09229.

\bibitem{Atomkiexpt} A. J. Krasznahorkay et al., Phys. Rev. Lett. 116, 042501 (2016). \\
doi:10.1103/PhysRevLett.116.042501; arXiv:1504.01527[nucl-ex].

\bibitem{Atomki2} A. J. Krasznahorkay et al., arXiv:2104.10075[nucl-ex].



\end{thebibliography}
\end{document}